%% file: potential.tex
\documentclass[aps,prb,reprint,showpacs,floatfix]{revtex4-1}
\usepackage{amsmath,graphicx,epsfig,color,verbatim,amsfonts,mathtools}
\usepackage{tikz}
\usepackage{array}
\usepackage{standalone}
\usetikzlibrary{positioning}

\pdfoutput=1

\let\oldhat\hat
\renewcommand{\vec}[1]{\mathbf{#1}}
\renewcommand{\hat}[1]{\oldhat{\mathbf{#1}}}

\newcommand{\rr}{\vec{R}}

\begin{document}
\title{A slave mode expansion for obtaining ab-initio interatomic potentials}
\author{Xinyuan Ai}
\email{xa2108@columbia.edu}
\affiliation{Department of Physics, Columbia University, New York, NY 10027}
\author{Yue Chen, Chris A. Marianetti}
\email{chris.marianetti@columbia.edu}
\affiliation{Department of Applied Physics and Applied Mathematics, Columbia University, New York, NY 10027}

\begin{abstract}
\noindent
Here we propose a new approach for performing a Taylor series expansion of the
first-principles computed energy of a crystal as a function of the nuclear
displacements.  We enlarge the dimensionality of the existing displacement
space and form new variables (ie. slave modes) which transform like
irreducible representations of the space group and satisfy homogeneity of free
space. Standard group theoretical techniques can then be applied to deduce the
non-zero expansion coefficients \emph{a priori}. At a given order, the
translation group can be used to contract the products and eliminate terms
which are not linearly independent, resulting in a final set of slave mode products.
While the expansion coefficients can be computed in a variety of ways,
we demonstrate that finite difference is effective up to fourth order. We
demonstrate the power of the method in the strongly anharmonic system PbTe.
All anharmonic terms within an octahedron are computed up to fourth order. A
proper unitary transformation demonstrates that the vast majority of the
anharmonicity can be attributed to just two terms, indicating that a minimal
model of phonon interactions is achievable. The ability to straightforwardly
generate polynomial potentials will allow precise simulations at length and
time scales which were previously unrealizable.
\end{abstract}

\date{\today}
\maketitle

\section{Introduction}
While the first-principles computation of the harmonic vibrational properties
of crystals with sufficient symmetry is ubiquitous\cite{Baroni2001515,Kunc1982406,Alfe20092622,Parlinski19974063,Walle200211}, the same cannot be said for
the anharmonic counterparts. The reasons for this are somewhat indirect.
Density functional theory (DFT), within the Born-Oppenheimer approximation, can
accurately predict the forces and stresses in many classes of materials and
therefore could be used to compute both quantum and classical dynamics of
the nuclei. However, the scaling of DFT severely restricts the applicability of
such a task to very short timescales and small unit cells. Generically, there
are a number of different approaches to overcoming this fundamental limitation
which exchange accuracy for efficiency, including fully empirical approaches
which replace DFT, semi-empirical electronic structure approaches\cite{Papaconstantopoulos2003413}, and linear
scaling DFT\cite{Bowler2012036503,Vandevondele20123565}. 

One obvious approach which has a long history is to perform a
Taylor series expansion of the energy as a function of the nuclear
displacements, allowing for extremely high precision up to some order and
within some range. While such an approach will have obvious limitation (ie.
large deformations, diffusion, etc.), it has a negligible computational cost
relative to DFT, allowing length and timescales which could not even be
considered within DFT. Furthermore, it has additional appeal in that the
expansion coefficients are basic materials properties. Understanding the
anharmonic interactions across a broad range of materials will help understand
a myriad of materials properties in terms of a low energy model. While the
number of anharmonic terms rapidly increases with the order of the expansion,
we demonstrate in this work that there is reason to be optimistic that a
minimal number of expansion coefficients can capture the bulk of the physics.
While the Hubbard and Anderson models have guided us for many years in terms of
understanding electronic phenomena in transition metal oxides and actinide
based materials\cite{Kotliar2006865}, analogues are clearly needed in the context of the interacting
phonon problem.

Some of the early executions of an anharmonic Taylor series expansion based on first-principles
calculations where executed by Vanderbilt et al in the context of Si\cite{Vanderbilt19895657} and by Rabe and Vanderbilt \emph{et al.} in the 
context of ferroelectric materials\cite{Kingsmith19945828,Zhong19941861,Zhong19956301}. These approaches
were quite successful, correctly capturing the proper ordering of different phases as a function of temperature and
even providing quantitatively accurate transition temperatures. In terms of the expansion, a variety of different philosophies
were taken in these works. The earliest of these works which focussed on Si\cite{Vanderbilt19895657} employed a quartic expansion in terms of 
bond bending and stretching variables 
in the spirit of earlier work of Keating\cite{Keating1966637,Keating1966674}. Coupling to strain becomes critical in the ferroelectric materials,
and an expansion similar to that of Pytte\cite{Pytte19723758} was used by Vanderbilt et al\cite{Kingsmith19945828} to encode the properties
of various perovskites. Subsequent work by Rabe et al\cite{Zhong19941861,Zhong19956301} utilized a novel lattice Wannier function approach\cite{Rabe199513236}
to perform an anharmonic expansion (see ref\cite{Iniguez20003127} for a related approach). 

With the continued explosion of computational resources, more recent works have
revisited this problem. Esfarjani and Stokes considered the generic Taylor
series expansion and all the symmetry constraints that the expansion must
satisfy\cite{Esfarjani2008144112}. They then generated a large data set from
first-principles calculations and fit the expansion parameters to the data under
the symmetry constraints.  
A number of materials and phenomena have been studied using this approach, including
the thermal conductivity in Si, half-Heusler compounds, and PbTe\cite{PhysRevB.84.085204, PhysRevB.84.104302, PhysRevB.85.155203}. 
Wojdel \emph{et al.}\cite{Wojdel2013305401} employed a different approach,
expanding in displacement differences between pairs of nuclei, similar in spirit to early model calculations\cite{Keating1966637,Keating1966674}, and they included point symmetry by
projecting displacement difference polynomials onto the identity
representation. Additionally,  Wojdel \emph{et al.}
explicity consider strain degrees of freedom and their coupling to local displacements, similar to 
earlier works in ferroelectric materials. 

It is also worth mentioning recent machine learning approaches that have the potential
to have significant impact in this space. Behler and Parrinello used a neural-network
to parameterize the DFT energy\cite{Behler2007146401}, and they have achieved impressive
results on Na\cite{Eshet2010184107,Eshet2012115701} and graphite/diamond\cite{Khaliullin2011693}.
These results suggest that appropriate neural-networks have the potential to accurately
describe structural phase transitions in a broad range of systems, though it is still unclear
if they have sufficient resolution to accurately capture phonons and higher derivatives
of the energy. Another approach in the context of machine learning is so-called compressive sensing,
which has been applied in the context of alloy theory to parameterize cluster expansions\cite{Nelson2013035125}
and has also shown promise in the context of lattice dynamics.

Despite the great successes of the aformentioned expansions, they have not yet
become ubiquitous, perhaps because it is nontrivial to execute the
parameterization. Here we introduce a new approach which combines many of the
advantages of the different methods discussed above. Our approach allows us to
circumvent the difficulties of fitting data across multiple orders, builds in
all the necessary symmetry from the beginning, is generally applicable,
provides a convenient notation to encode our parameters such that others may
use them, and in the case of PbTe we show that a physically motivated unitary
transformation can compress hundreds of anharmonic terms into just two.

\section{Method}
\subsection{Background}

We will start by considering the total energy of a crystal assuming that the Born-Oppenheimer approximation is valid. The Taylor series expansion of the total energy as a function of
the nuclear displacements can be written as follows\cite{Esfarjani2008144112}:

\begin{equation}
\begin{aligned}
V&=\sum_{\alpha\beta\rr_a\rr_b} \Psi(\rr_a,\rr_b)_{\alpha\beta}u^{\rr_a}_{\alpha}u^{\rr_b}_{\beta}\\
&+\sum_{\alpha\beta\gamma\vec{R}_a\vec{R}_b\vec{R}_c}
\Psi(\rr_a,\rr_b,\rr_c)_{\alpha\beta\gamma} u^{\vec{R}_a}_{\alpha} u^{\vec{R}_b}_{\beta} u^{\vec{R}_c}_{\gamma} \\
&+ \sum_{\substack{\alpha\beta\gamma\delta \\ \vec{R}_a\vec{R}_b\vec{R}_c\vec{R}_d}}
\Psi(\vec{R}_a,\vec{R}_b,\vec{R}_c,\vec{R}_d)_{\alpha\beta\gamma\delta}
u^{\vec{R}_a}_{\alpha} u^{\vec{R}_b}_{\beta} u^{\vec{R}_c}_{\gamma} u^{\vec{R}_d}_\delta\\
&+\cdots
\end{aligned}
\label{rhamed}
\end{equation}
Where $\Psi$ are the direct expansion coefficients, $u$ are the nuclear displacements, $\rr=n_1\vec{v}_1+n_2\vec{v}_2+n_3\vec{v}_3$ ($n_i$
are integers, $\vec{v}_i$ are unit cell vectors), $\alpha,\beta,\gamma,\delta$ label both the displacement direction
(ie. $x,y,z$) and the atom within the unit cell. 
The number of terms dramatically increases as the
order increases. Therefore, a
condition for this expansion to be useful is locality: the expansion coefficients must decay
sufficiently rapidly in some representation for terms beyond quadratic order. This cannot be known
\emph{a priori} and only explicit testing could determine the viability of this
approach. Symmetry will be crucial both to reduce the number of terms at a given
order and to ensure that the expansion is robust for use in simulations. The following
symmetries must be satisfied:
\begin{enumerate}
\item The energy must be invariant to all space group operations.
\item  Homogeneity of free space with respect to rigid translation. If the entire crystal is shifted by an arbitrary constant, there cannot be any change in the total energy nor its derivatives. 
\item  Homogeneity of free space with respect to rigid rotation. If the entire crystal is rotated about some point by an arbitrary amount, there cannot be any change in the total energy nor its derivatives. 
\item If the energy function is analytic, the derivatives will be invariant of the order in which they are taken.
\end{enumerate}
These symmetries result in a series of constraints on the expansion coefficients\cite{Esfarjani2008144112}.
The central task at hand is to actually compute the derivatives of the energy
with respect to the atomic displacements and ensure that they satisfy all of
the symmetries.

\subsection{Slave Mode Expansion}
Executing the expansion would be far more straightforward 
if the symmetry could be somehow imposed from the beginning. This can be
achieved at the expense of enlarging the dimensionality of the system. Instead
of using nuclear displacement parameters $u$, we will introduce  so-called
\emph{slave modes} $\phi$ which transform like irreducible representations
of the space group and satisfy homogeneity of free-space. These slave modes
may then be used to expand the potential, and all of the symmetry constraints
will be built into the expansion. Enlarging the dimensionality does come at an expense, as
some of the slave mode products will be linearly dependent, but this can be handled in a 
straightforward fashion. 
It should be noted that the one symmetry we do not consider
is homogeneity of free space with respect to rotations, which will link coefficients at different orders\cite{Esfarjani2008144112}. Fortunately,
any associated errors will not accumulate due to the satisfaction of point group symmetry.
We now expand the energy in terms of the slave modes:
\begin{equation}
\begin{aligned}
V&=\sum_{\rr s} \sum_{\alpha i} 
\Phi_{\alpha}^{s}\hspace{1mm}
\phi_{\alpha\rr s}^{(i)} \phi_{\alpha\rr s}^{(i)}  \\
&+\sum_{\rr s} \sum_{\substack{\alpha\beta\gamma\\ \xi,ijk}}
\Phi_{\alpha\beta\gamma}^{s\xi}\Theta_{\alpha\beta\gamma}^{\xi,ijk}\hspace{1mm}
\phi_{\alpha\rr s}^{(i)} \phi_{\beta\rr s}^{(j)} \phi_{\gamma\rr s}^{(k)} \\
&+ \sum_{\rr s} \sum_{\substack{\alpha\beta\gamma\delta \\ \xi, ijkl}}
\Phi_{\alpha\beta\gamma\delta}^{s\xi}\Theta_{\alpha\beta\gamma\delta}^{\xi,ijkl}\hspace{1mm}
\phi^{(i)}_{\alpha\rr s} \phi^{(j)}_{\beta\rr s} \phi^{(k)}_{\gamma\rr s} \phi^{(l)}_{\delta\rr s} +\cdots 
\end{aligned}
\label{rhamed}
\end{equation}
where $\alpha,\beta,\gamma,\delta$ label irreducible representations, $i,j,k,l$ label rows of
a given irreducible representation, $\xi$ labels a given identity representation within the 
product representation, $\rr$ is a lattice vector, $s$ labels a cluster within the unit cell, $\Theta$ are the Clebsch-Gordan (CG) coefficients,
$\Phi$ are the irreducible expansion coefficients, and $\phi$ are the slave modes. 
It should be noted that cross terms between the clusters with different $\rr$ or $s$ are not written as
their contribution can be accounted for by simply including larger clusters.
The CG coefficients are a group theoretical construct which are
independent of any particular application, and these may be straightforwardly computed. However, care
must be taken to ensure that a consistent phase convention has been used as there is no unique definition. The slave
clusters $\phi$ are a linear combination of atomic displacements which transform like
the irreducible representation of a given point group in the crystal. While we have explicitly written out
the quadratic terms, we will assume that these will normally be obtained using traditional approaches to
compute phonons. 

There is a wide degree of flexibility in choosing the slave modes, and the optimum choice may depend on 
the material and the use of the method. Here we will outline a typical scenario, and specific cases will be
dealt with later in the manuscript.

\begin{enumerate}
\item  Determine a cluster of atoms for which the anharmonic terms will be
included. This cluster will be associated with a given unit cell (typically primitive), though it could contain
atoms which are outside of the unit cell. As
the size of the cluster increases, the number of terms in the expansion will
increase markedly, so this choice must be made judiciously.  \emph{At least two
atoms must be present in this cluster}. We will refer to this as the slave cluster.

\item A center of highest symmetry should be identified for the chosen cluster
and the associated point group should be determined. Each atom in the cluster
will have $d$ degrees of freedom, where $d$ is the dimension of space.
The displacement vectors should then be projected onto the
irreducible representation of the point group. 

\item $d$ of the irreducible representations that correspond to a uniform shift
of the cluster need to be eliminated as they would violate homogeneity of free
space. The  remaining modes are the \emph{slave modes}, and   they can
essentially be thought of as molecular entities. The  usual methods of finite
group theory may be used to show that only products transforming like the
identity are non-zero\cite{Cornwell,Tinkham}.

\item All non-translation space group operations should be used to determine if
translationally inequivalent slave clusters are generated from the initial set. 

\item The translation group may then be used to generate all translationally
equivalent sets.
\end{enumerate}
It may be useful to have multiple types of slave clusters associated with each unit cell, and
then the above procedure will be executed for each slave cluster. This will
indeed be the case for PbTe.

At this point, one has an expansion which respects all of the necessary
symmetries, albeit at the expense of increasing the dimensionality of the
system. If one began with a crystal having $\alpha$ atoms per unit cell in $d$ dimensions and there are
$N$ unit cells in the crystal, then the total number of degrees of freedom would be $(N\alpha  - 1 )d $. 
If one chose a single cluster per unit cell having $z$ atoms, then the total number of degrees of freedom
would be $(z-1)dN$. Under normal conditions $z> 2$, and therefore the dimensionality of the system has increased. 
This does give rise to several issues. First, if one wanted to use the slave modes as independent variables,
then a constraint would have to be satisfied in order to be sure that the vibrational state is physical. In other words,
an arbitrary vector in the space of slave modes will not necessarily have a corresponding vector in the space of
displacements. However, this poses no problem in this work as we will always be using the slave modes as dependent variables. 
The second issue is that slave mode products at a given order which are irreducible with respect to point symmetry will not necessarily be linearly independent when lattice
translations are included, and therefore certain mode products must be eliminated to remove linear dependency.
This is a penalty that must be dealt with in order to take this approach. A final point worth noting is that slave modes on different
sites are not orthogonal as we have presented them in this work. This means that an amplitude for a slave mode on one-site
will induce a non-zero amplitude on its neighbor. However, this poses no real challenges to the method.

\subsection{Slave mode Expansion for the $2 d$ Square Lattice}
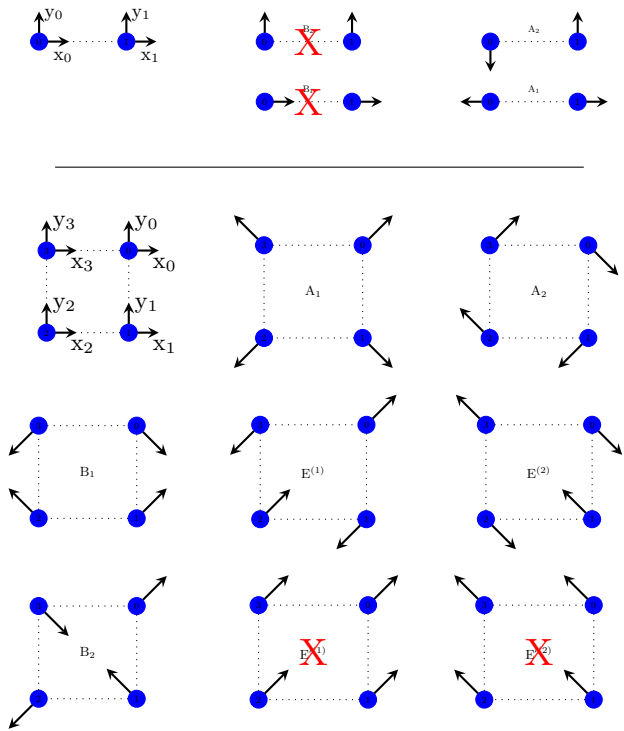
\begin{figure}
\begin{center}

\input{fig_dimer_modes.tex}
\line(1,0){200}
\vspace{5mm}

\input{fig_square_modes.tex}
  \caption{(Top panel) Normal modes for the dimer cluster in the square lattice. (Bottom panel) Normal modes for the square
  cluster in the square lattice. }
  \label{square_modes}
  \end{center}
\end{figure}

To illustrate the slave mode expansion we apply it to a two
dimensional square lattice with one atom in one unit cell. We will explore two
difference choices for slave modes. First, let us consider a cluster of two
nearest-neighbor atoms (ie. dimer cluster). In this case, we will choose the center of the cluster
as the midpoint of the bond (see figure \ref{square_modes} top panel), which
will have point group $C_{2v}$. The $C_{2v}$ point group allows four different
irreducible representations, and we follow the usual convention of
$A_1,A_2,B_1,B_2$\cite{Cornwell}.  The representation for the dimer cluster is four dimensional
and can be decomposed as $\Gamma=A_1\oplus A_2\oplus B_1\oplus B_2$ (see figure \ref{square_modes} top panel).
The two normal modes $B_1\oplus B_2$ correspond to uniform shifts of the
cluster, and therefore these modes will be removed, as indicated by the red
$X$, leaving only $A_1\oplus A_2$.  Using the Great
Orthogonality Theorem (GOT)\cite{Cornwell,Tinkham}, one can deduce that only the products that
transform like the identity will be nonzero, and these can be determined by
inspecting the character table of $C_{2v}$.  There will be two slave mode products at second
order: $\phi_{A_1}^2$ and $\phi_{A_2}^2$. At third order there will be two
terms: $\phi_{A_1}^3$ and $\phi_{A_2}^2\phi_{A_1}$. At fourth order there will
be three terms: $\phi_{A_1}^4$, $\phi_{A_2}^4$, and $\phi_{A_2}^2\phi_{A_1}^2$.
One can easily proceed to higher orders, but we will remain at quadratic order for the sake of simplicity in this example. At this point, one needs to see
if any non-translational symmetry elements will generate 
a new slave mode center which is
translationally inequivalent. Clearly, the $C_{4v}$ group at the
center of the square will rotate the dimer from a horizontal one to a
vertical one. The rotated slave mode products will have the identical coefficients. The next step
would be to use the translation group to determine if any set of slave mode products are linearly
dependent. This can straightforwardly be checked by summing over all modes that have overlap,
expanding the slave mode products into displacement products, and determining the rank of the resulting
polynomial matrix. 
In this simple case there will be no linear dependence because dimers only share edges, but this will not
be the case when we treat a larger cluster below.

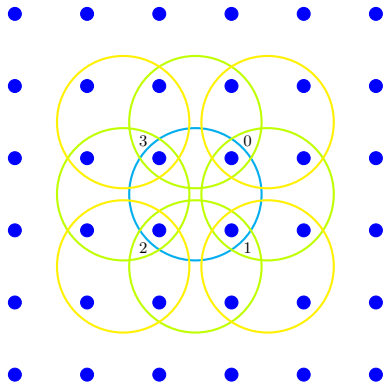
\begin{figure}
\begin{center}

\input{square_lattice_overlap.tex}
  \caption{ A schematic illustrating the summation over square slave modes which overlap with the central cluster. Eight neighboring slave
  clusters must be summed over.}
  \label{sum_modes}
  \end{center}
\end{figure}

The second illustration would be to consider interactions within a square, and
this will be carried to second order. In this case the point symmetry group will be
$C_{4v}$, which allows for five irreducible representations: $A_1, A_2, B_1, B_2,
E$\cite{Cornwell}.  The square cluster representation is eight dimensional, and these can be
decomposed as $\Gamma=A_1\oplus A_2 \oplus B_1 \oplus B_2 \oplus 2E$ (see figure
\ref{square_modes} bottom panel).  In this case the $E$ irreducible
representation appears twice. One set of the $E$ irreducible representations
can be chosen to be shifts of the cluster while the other set will be obtained via orthogonalization.
The $E$ representation corresponding to a shift will be removed, and the slave mode representation will be
$A_1\oplus A_2 \oplus B_1 \oplus B_2 \oplus E$. In this case, there are
no nontranslational symmetry elements that will generate translational
inequivalent slave clusters. At second order there will be the following products: $\phi_{A_1}^2, \phi_{B_1}^2, \phi_{A_2}^2, \phi_{B_2}^2, \phi_{E^{(1)}}^2+\phi_{E^{(2)}}^2$.
Finally, the translation group must be used to check for linear dependence. 
One can check for linear dependency by summing all slave mode products that overlap a given cluster (see
figure \ref{sum_modes} for an illustration), multiplying out the slave mode products into displacement polynomials,
and representing them in the space of displacement polynomials as follows:
\begin{align}\label{rank}
\frac{1}{4}\left( \begin{array}{cccccccc}
2 &   2 & - 2 & - 1 & - 1 &   2 & - 1  & \dots \\
2 &   2 & - 2 & - 1 &   1 &   2 & - 1  & \dots \\
2 & - 2 &   2 & - 1 &   1 &   2 & - 1  & \dots \\
2 & - 2 &   2 & - 1 & - 1 &   2 & - 1  & \dots \\
4 & - 4 & - 4 &   2  &  0 &   4 &   2  & \dots 
\end{array}\right)
\left( \begin{array}{c}
x_2^2 \\ x_2x_3 \\ x_2x_1 \\ x_2x_0 \\ x_2y_0 \\  x_3^2 \\ x_3x_1 \\ \vdots
\end{array}\right)
\end{align}

The rank of the resulting matrix is 4, and one can show that 
one of the products $\phi_{A_1}^2, \phi_{B_1}^2, \phi_{A_2}^2, \phi_{B_2}^2$
must be removed.
Therefore, there are four expansion coefficients corresponding to the following products: $\phi_{A_1}^2, \phi_{B_1}^2,  \phi_{B_2}^2, \phi_{E^{(1)}}^2+\phi_{E^{(2)}}^2$.
Typically, one will actually compute the direct expansion coefficients $\Psi$ using DFT,
and therefore we will need to relate the slave mode product coefficients $\Phi$ to $\Psi$.
At a given order, this can simply be written as a matrix equation, and we illustrate this
at second order for this scenario: 
\begin{align}\label{direct_to_slave}
\frac{1}{4}\left( \begin{array}{cccc}
2  &  2 &  2 &  4 \\
2  &  2 & -2 & -4 \\
-2 & -2 &  2 & -4 \\
-1 & -1 & -1 &  2 \\
-1 &  1 & -1 &  0 \\
2  &  2 &  2 &  4 \\
-1 & -1 & -1 &  2 \\
\vdots &  \vdots  &  \vdots  &  \vdots 
\end{array}\right)
\left( \begin{array}{c}
\Phi_{A_1} \\ \Phi_{B_1} \\ \Phi_{B_2} \\ \Phi_{E} 
\end{array}\right)=
\left( \begin{array}{c}
\Psi_{x_2x_2} \\ \Psi_{x_2x_3} \\ \Psi_{x_2x_1} \\ \Psi_{x_2x_0} \\ \Psi_{x_2y_0} \\  \Psi_{x_3x_3} \\ \Psi_{x_3x_1} \\ \vdots
\end{array}\right)
\end{align}
One needs to compute enough direct expansion coefficients such that the number of rows is greater than or equal to the number
of columns. If the DFT computations had no imprecisions, one could simply compute as many direct coefficients as slave coefficients,
but it is far more robust to create an overdetermined scenario. It is important to note that the above relation is
only robust if sufficiently large slave modes are chosen such that they have sufficiently decayed with respects to distance.

\section{Slave Mode Expansion for Rock-Salt: PbTe}
\begin{figure}[htb]
\begin{center}
\includegraphics[width=\linewidth,clip= ]{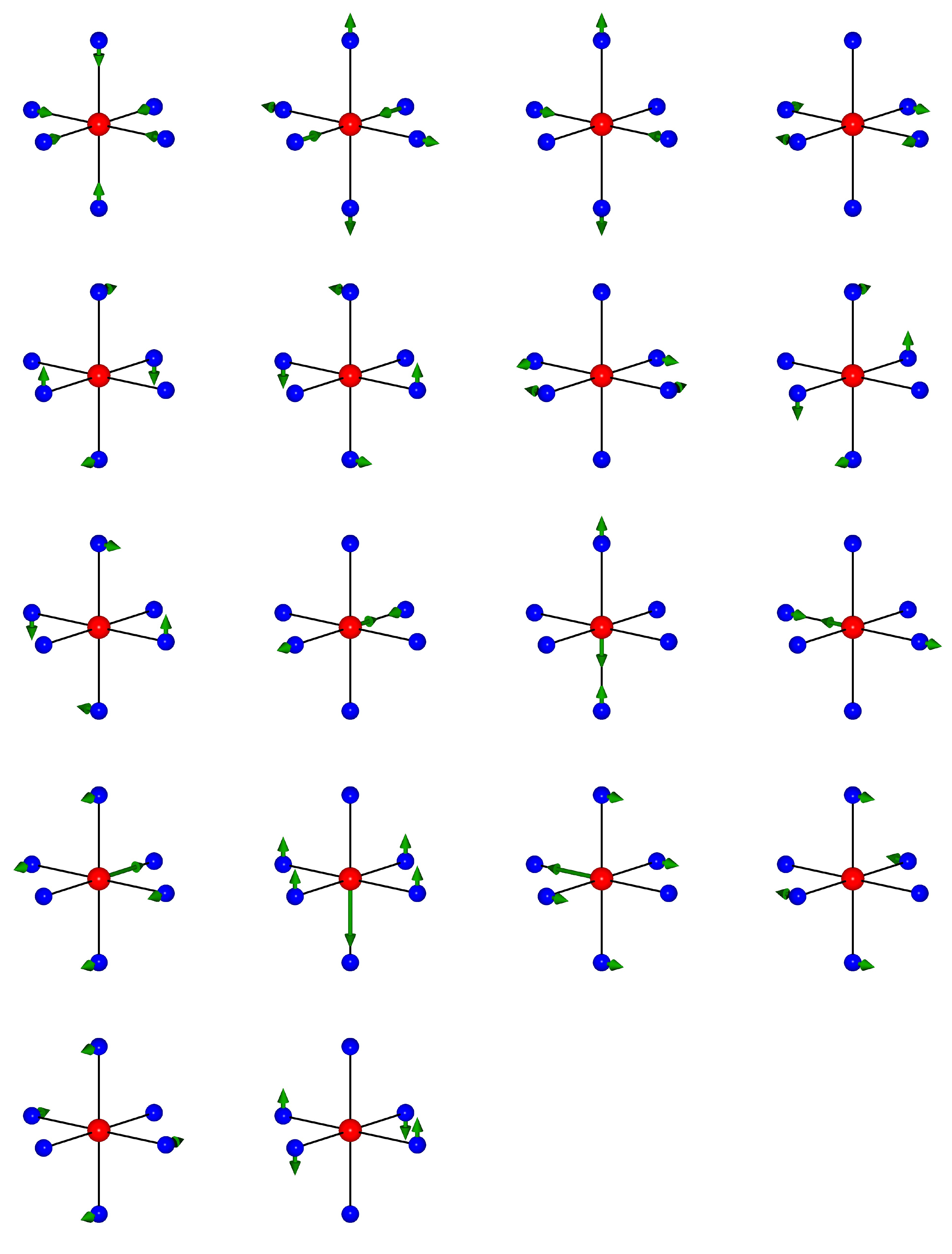}
\end{center}
\caption{Octahedral modes transforming as the irreducible representations of the point group. The three $T_{1u}$ modes which shift 
the octahedron have been removed. Reading from left to right and top to bottom, the modes are $A_{1g}, E_g, T_{1g}, T_{2g}, 2T_{1u}, \textrm{and } T_{2u}$.}
\label{octmodes}
\end{figure}

\begin{figure}[htb]
\begin{center}
\includegraphics[width=\linewidth,clip= ]{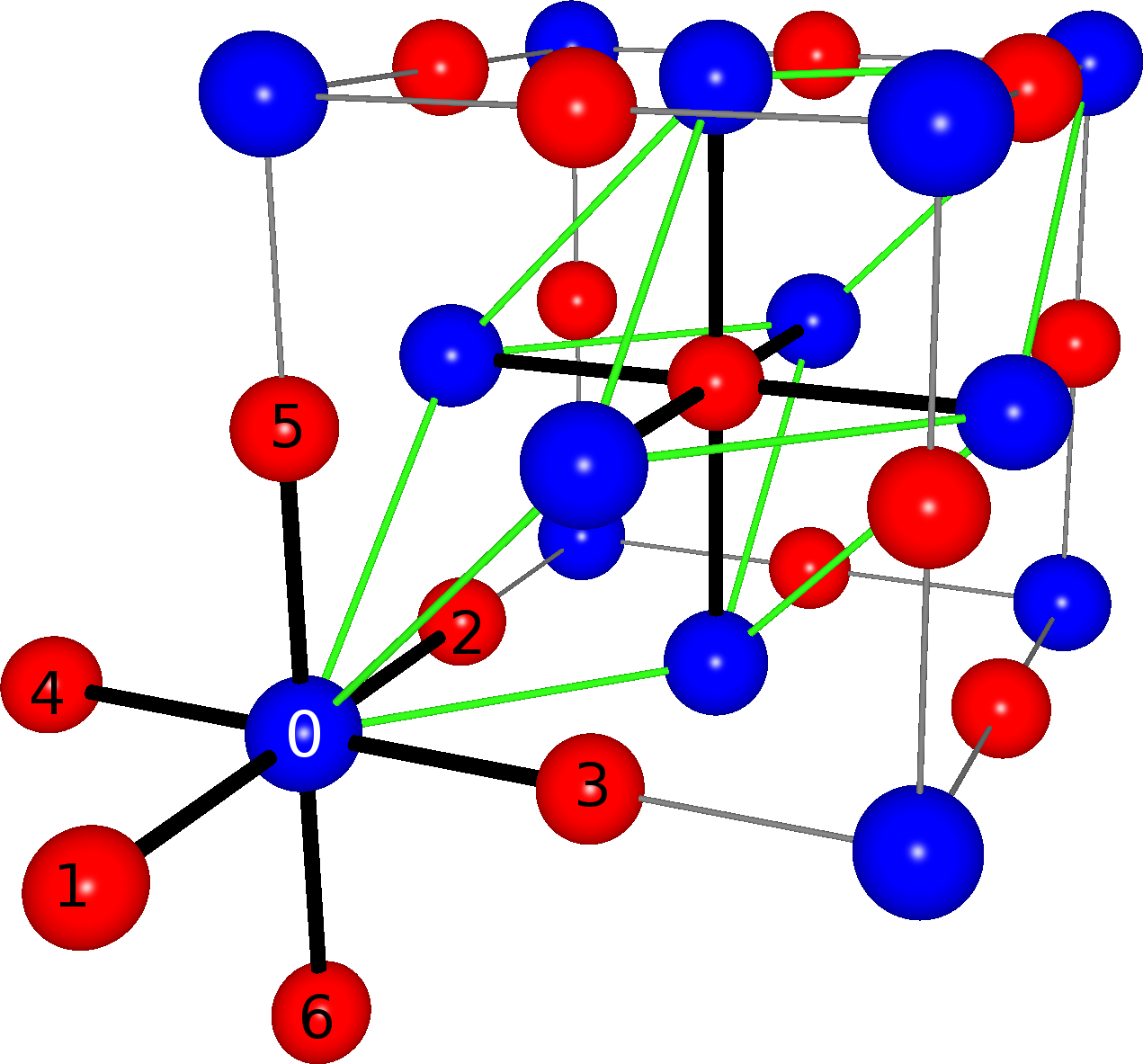}
\end{center}
  \caption{ A section of the rock salt structure. The primitive unit cell is given in green. The two slave clusters associated with the primitive
  unit cell are denoted by atoms connected with bold lines. The octahedral numbering convention is shown. }
  \label{rocksalt}
\end{figure}

Here we apply the slave mode expansion to the rock-salt structure of PbTe. 
We will choose a primitive unit cell having vectors ${\bf a_1}=a/2(1,1,0)$, ${\bf a_2}=a/2(0,1,1)$, and ${\bf a_3}=a/2(1,0,1)$,
with a Pb atom at $(0,0,0)$ and a Te atom at $(\frac{1}{2},\frac{1}{2},\frac{1}{2})$ (fractional coordinates, see figure \ref{rocksalt}).
The
first task is to pick the clusters within which we will retain terms beyond
quadratic. There are two natural choices: the Pb-Te dimer and the octahedron (both Pb centered and Te centered). We will
begin by considering the octahedra as the cluster of choice (see section \ref{minmod} for the dimer), which implies that we will have anharmonic terms within next
nearest neighbor for both Pb and Te. There will be two slave clusters associated with each primitive unit cell, each having $O_h$ point symmetry, and these correspond to atoms
connected with bold black lines in figure \ref{rocksalt}. Translationally equivalent clusters can be generated by shifting with the
primitive lattice vectors (denoted as green lines in figure \ref{rocksalt}). We now proceed to decompose the displacement vectors
into irreducible representations of the $O_h$ point group (see figure \ref{rocksalt} for octahedral labeling convention), and these
are listed in figure \ref{octmodes} to define the phase conventions which we choose.

The octahedral slave representation can be decomposed into $\Gamma=A_{1g} \oplus E_g \oplus T_{1g} \oplus  T_{2g} \oplus 2T_{1u} \oplus T_{2u}$, where we have
remove a $T_{1u}$ manifold which rigidly shifts the octahedron. One can then form product representations in a given octahedron, showing that there
are 29 unique products at third order and 153 unique products at fourth order. This will be the case for both Pb and Te centered octahedron. 
Nontranslational
symmetry elements will not generate any translationally inequivalent slave clusters. Employing the translation group, one can demonstrate that 
some of the terms are redundant. In particular, two terms will be removed at third order, and four terms will be removed at fourth order. The final result
is that there are 56 terms at third order and 302 terms at fourth order, for a total of 358 terms up to fourth order and within next-nearest neighbor range.

The third order products and corresponding coefficients are listed in table \ref{term3t}, while fourth order terms are listed in table \ref{term4t}.
It should be noted that within a given product some of the identity representations are redundant or zero due to the fact that we are dealing
with self-products. For example, $E_g \otimes E_g \otimes E_g \otimes E_g=5E_g\oplus 3A_{2g}\oplus 3A_{1g}$, but only one of the three $A_{1g}$
representations is unique when projecting the displacement product vector.
It is important to note the phase convention we chose in constructing the Clebsch-Gordan coefficients. The vectors in each product subspace are labeled
from $1\dots N$ when taking the following ordering: 
\begin{align}\label{}
|1,0,0,0,0,\cdots \rangle ,  |0,1,0,0,0,\cdots  \rangle ,   \cdots
\end{align}
In tables \ref{term3t} and \ref{term4t} we list the vector which was used to project onto the identity, and this sets the phase 
convention for our Clebsch-Gordan coefficients wich can straightforwardly be constructed.

\begin{center}
\begin{table}
\footnotesize
\scalebox{0.75}{
  \begin{tabular}{| l | l | l | l |}
  \hline
   { \bf Product } & {\bf Phase} & {\bf Pb-centered $\Phi$ } & { \bf Te-centered $\Phi$} 

\input{table_coeff3.tex}

 \\ \hline

  \end{tabular}
        }
  \caption{Nonzero third order products and the corresponding expansion coefficients. The second column lists which product vector was used to 
  project the identity and create the Clebsch-Gordan coefficients for each corresponding coefficient $\Phi$. Terms designated N/A were those
  removed by the translation group.}
  \label{term3t} 
\end{table}
\end{center}

\section{Computing expansion coefficients for PbTe}
Having determined the slave mode expansion up to 4$^{th}$ order and within next-nearest neighbor,
the slave mode coefficients must be computed. In general, there are be many approaches to execute
this task. Firstmost, as described above, we will assume that the harmomic terms have been computed using
tranditional approaches for computing phonons from first-principles, such as density functional perturbation theory\cite{Baroni2001515}
or finite displacement supercell approaches\cite{Alfe20092622,Kunc1982406}. Therefore, we are only concerned with computing the
third and fourth order terms. An obvious approach would be to construct a large data set of atomic displacements
in the anharmonic regime and compute the corresponding energies using DFT. This dataset may then be used to fit
the slave mode expansion coefficients using standard procedures. The drawback of such an approach is that
one is always faced with the problems of overfitting or incuding data which which is beyond fourth order. While
there are standard statistical methods to address such problems, we believe other approaches are likely more
straightforward. Another approach would be to  compute
individual expansion coefficients in the direct expansion (ie. equation \ref{rhamed}), analagous to what is
done for the harmonic case in phonons. One could either use the $2N+1$ theorem from density functional perturbation
theory\cite{Gonze198913120,Debernardi19951819,Deinzer2003144304}, or a supercell approach using finite displacements could be used. We will opt for the latter in this work.

The computed direct expansion terms are only of limited use given that small errors within the numerical
implementation of DFT will prevent the computed direct terms from satisfying all the necessary symmetries.
However, there is a linear relation between the slave mode coefficients and the direct expansion coefficients (see equation \ref{direct_to_slave} for an example).
Therefore, one simply needs to compute enough direct coefficients such that the slave mode coefficients are
uniquely defined. In the case of PbTe, we will need to compute at least 56 direct coefficients at third order
and 302 at fourth order. In practice, it is desirable to compute more than the minimum number to minimize
the effects of error within the DFT finite difference calculations.

These linear relations will properly average out any small noise from the direct coefficients
and enforce all symmetry relations.  What should be apparent is that these
relations assume a truncation in the range of the slave modes. This is clearly
an approximation which relies on a sufficient degree of locality in order to be
accurate, and we will show that our truncation of an octahdron for PbTe is reasonable. The
other major potential source of error is the convergence of the direct finite
difference terms which will be dealt with below. While it would be desirable to
directly compute the slave mode coefficients, this is not straightforward
as the slave modes are not orthogonal.

\subsection{DFT runs and Finite Difference}
As outlined above, the direct expansion coefficient will be computed with  finite difference. Given that the forces are known from the Hellman-Feynman Theorem\cite{Martin2008}, the first derivatives will all
be known for a given DFT computation. Using a central finite difference, a derivative containing up to four variables can generically be written:
\begin{align} 
\nonumber&\frac{\partial^n E}{\partial q_\alpha^h \partial q_\beta^i \partial q_\gamma^j \partial q_\delta^k}=
          \frac{\partial^{n-1} F_\alpha}{\partial q_\alpha^{h-1} \partial q_\beta^i \partial q_\gamma^j \partial q_\delta^k} \\
\nonumber&\approx \frac{1}{2\Delta}\frac{\partial^{n-2}}{\partial q_\alpha^{h-1} \partial q_\beta^{i-1} \partial q_\gamma^j \partial q_\delta^k} \left[F_\alpha(q_\beta + \Delta)-F_\alpha(q_\beta - \Delta)\right]\\
\nonumber&\approx \frac{1}{4\Delta^2} \frac{\partial^{n-3}}{\partial q_\alpha^{h-1} \partial q_\beta^{i-1} \partial q_\gamma^{j-1}\partial q_\delta^k} 
\left[
F_\alpha(q_\beta + \Delta,q_\gamma+\Delta)- \right. \\
\nonumber& \hspace{5mm} F_\alpha(q_\beta - \Delta,q_\gamma+\Delta) - F_\alpha(q_\beta + \Delta,q_\gamma-\Delta) \\
\nonumber& \hspace{5mm} \left. +F_\alpha(q_\beta - \Delta,q_\gamma-\Delta) \right] \\
\nonumber &\approx \dots \\
\nonumber & =\frac{1}{{(2\Delta)}^{n-1}}
 \sum_{n_\alpha = 0}^{h-1} \sum_{n_\beta = 0}^i \sum_{n_\gamma = 0}^j
 \sum_{n_\delta = 0}^k 
 {h-1 \choose n_\alpha } {i \choose n_\beta } {j \choose n_\gamma } {k \choose n_\delta } \\
 \nonumber & (-1)^{n_\alpha+n_\beta+n_\gamma + n_\delta} F_\alpha(q_\alpha+(h-1-2n_\alpha)\Delta,
 q_\beta+(i-2n_\beta)\Delta, \\
 & q_\gamma+(j-2n_\gamma)\Delta,
  q_\delta+(k-2n_\delta)\Delta)
\label{fd}
\end{align}
with $\alpha,\beta,\gamma,\delta$ label both the atom and the displacement vector,
$n=h+i+j+k$ which label the order of the derivative, $F$ is force, and $\Delta$
is the finite difference displacement. For a third order term, four DFT
computations will be needed, while eight will be needed for a fourth order
term.

The forces are computed within the framework of Density Functional Theroy which
is carried out using the generalized gradient approximation (GGA) by Perdew and
Wang\cite{PhysRevLett.77.3865} as implemented in the Vienna ab initio
simulation package (VASP) \cite{Kresse1993558,Kresse199414251,Kresse199615,Kresse199611169,Kresse19991758}. Gamma centered k-meshes
depending on the supercell size are applied and a $3\times3\times3$ mesh is used
for the smallest 64-atom supercell. Charge self-consistency is performed until
the energy is converged to within $10^{-5}$ eV, and a plane wave cutoff of $175-350eV$ was used depending
on the particular computation. Spin-orbit coupling was not utilized.

\begin{figure}[htb]
\begin{center}
\includegraphics[width=\linewidth,clip= ]{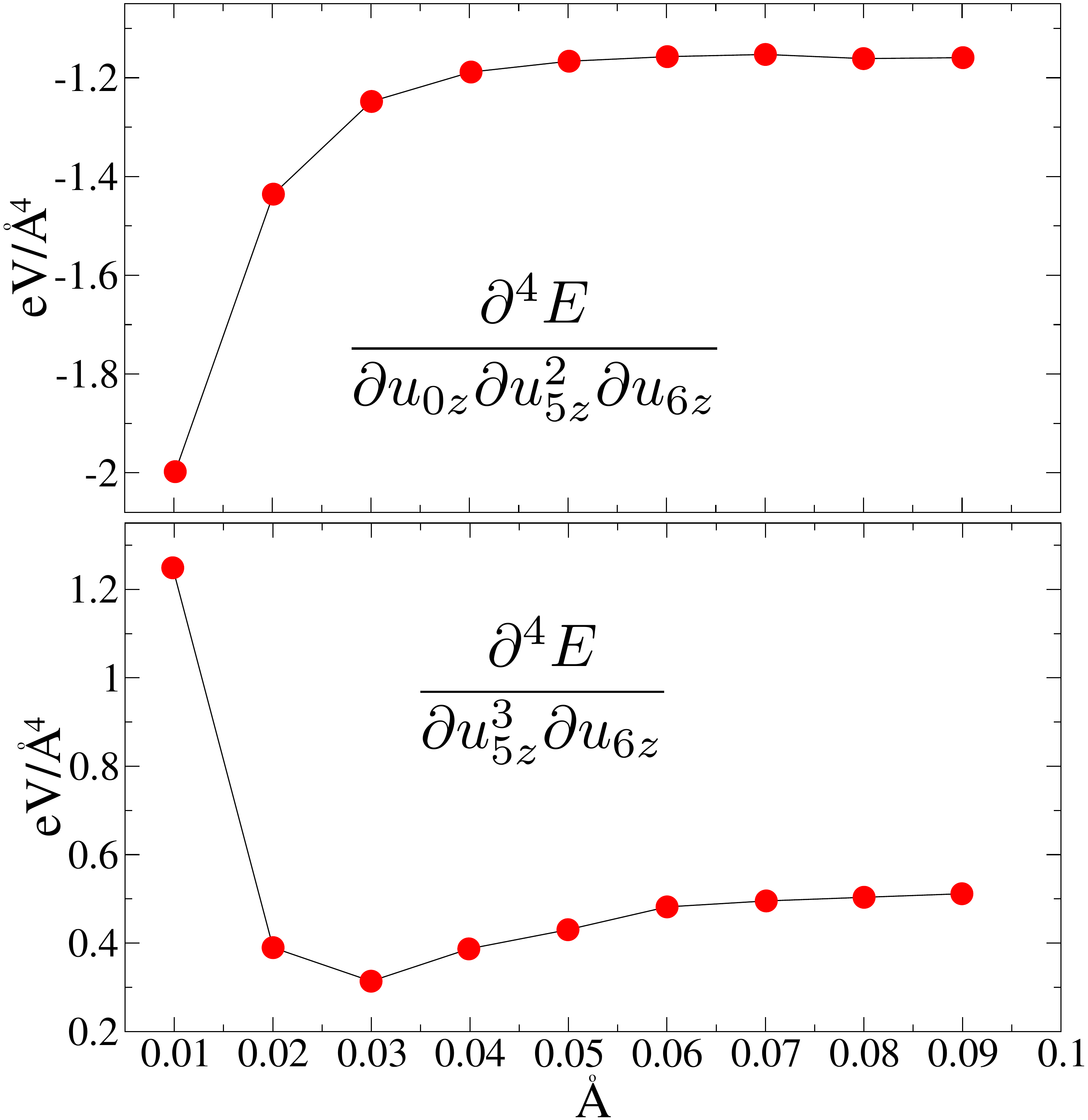}
\end{center}
  \caption{Fourth order derivatives computed using central step finite difference as a function of $\Delta$ for a conventional supercell choice of $2\times 2\times 3$ (ie. 96 atoms). }
  \label{delta_conv}
\end{figure}

\begin{figure}[htb]
\begin{center}
\includegraphics[width=\linewidth,clip= ]{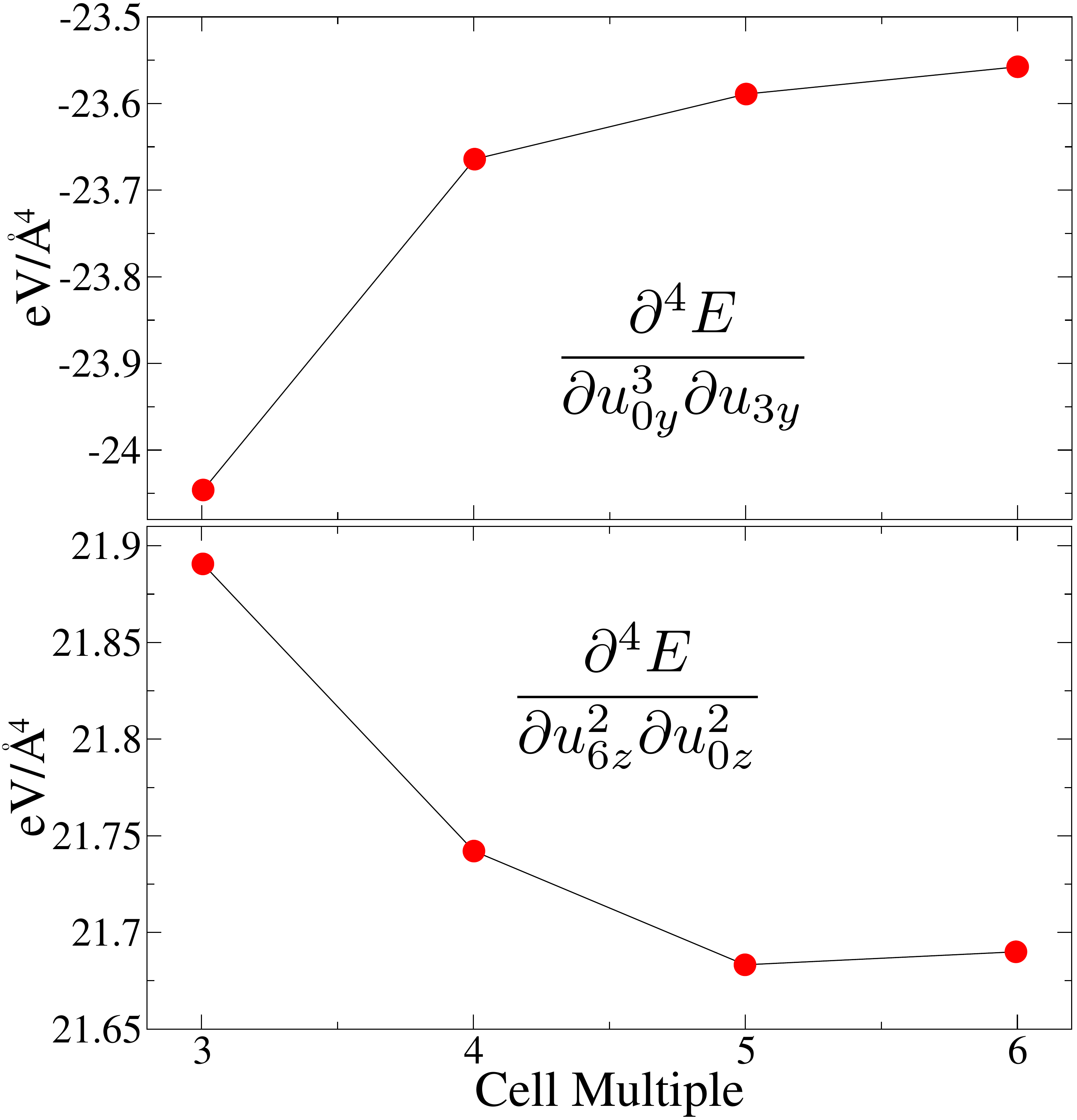}
\end{center}
\caption{Fourth order derivatives computed using central step finite difference as a function of conventional supercell size in the $y$-direction for $\Delta=0.07\AA$}
\label{cellsize}
\end{figure}

In order to be sure the direct coefficients are robustly computed within finite difference, one must test
for convergence with respect to the displacement size $\Delta$ in addition to the supercell size. 
If $\Delta$ is chosen to be too small, a probibitive planewave cuttoff and k-point mesh will be required, while if it is too large
higher order terms will taint the computation. Therefore, there will be an optimum $\Delta$ which will be both efficient and accurate,
and this will strongly depend on the order of the derivative. In order to illustrate this point, the values of two different fourth order
expansion coefficients are plotted as a function of $\Delta$ (see figure \ref{delta_conv}). A clear platuea emerges in both cases,
revealing a robust value for $\Delta$. After examining a wide range of different types of direct coefficients, we found that
$\Delta=0.01\AA$ is reliable for third order while $\Delta=0.07\AA$ is reliable for fourth order. Supercell size must also be studied
to be sure that images are not interacting with one another. The minimum supercell dimension that was used was twice the conventional (ie. cubic)
cell size, while the maximum was six times the conventional cell size. In order to illustrate this, we plot two fourth order coeefficients as
a function of unit cell size along a particular dimension (see figure \ref{cellsize}), demonstrating that the changes in the coefficients
are diminishing with increasing cell size. Our convergence criteria for supercell dimension was determined based on the largest finite difference
coefficient at a specific order, and for third order the unit cell size was increased until changes were within 0.01 $eV/\AA^3$ while the
threshold was 0.1$eV/\AA^4$ for fourth order.

\begin{center}
\begin{table*}
\footnotesize
\tabcolsep=0.11cm
\scalebox{0.75}{
        \begin{tabular}{| l | m{1.25in} | m{2.25in} | m{2.25in} |}
        \hline
   { \bf Product } & {\bf Phase} & {\bf Pb-centered $\Phi$ } & { \bf Te-centered $\Phi$}   \\ \hline

\input{table_coeff4_withseeds.tex}

 \\ \hline

        \end{tabular}
        }
  \caption{Nonzero fourth order products and the corresponding expansion coefficients. The second column lists which product vector was used to 
  project the identity and create the Clebsch-Gordan coefficients for each corresponding coefficient $\Phi$. }
        \label{term4t} 
\end{table*}
\end{center}

\subsection{Slave mode expansion coeefficients}
\begin{figure}[htb]
\begin{center}
\includegraphics[width=\linewidth,clip= ]{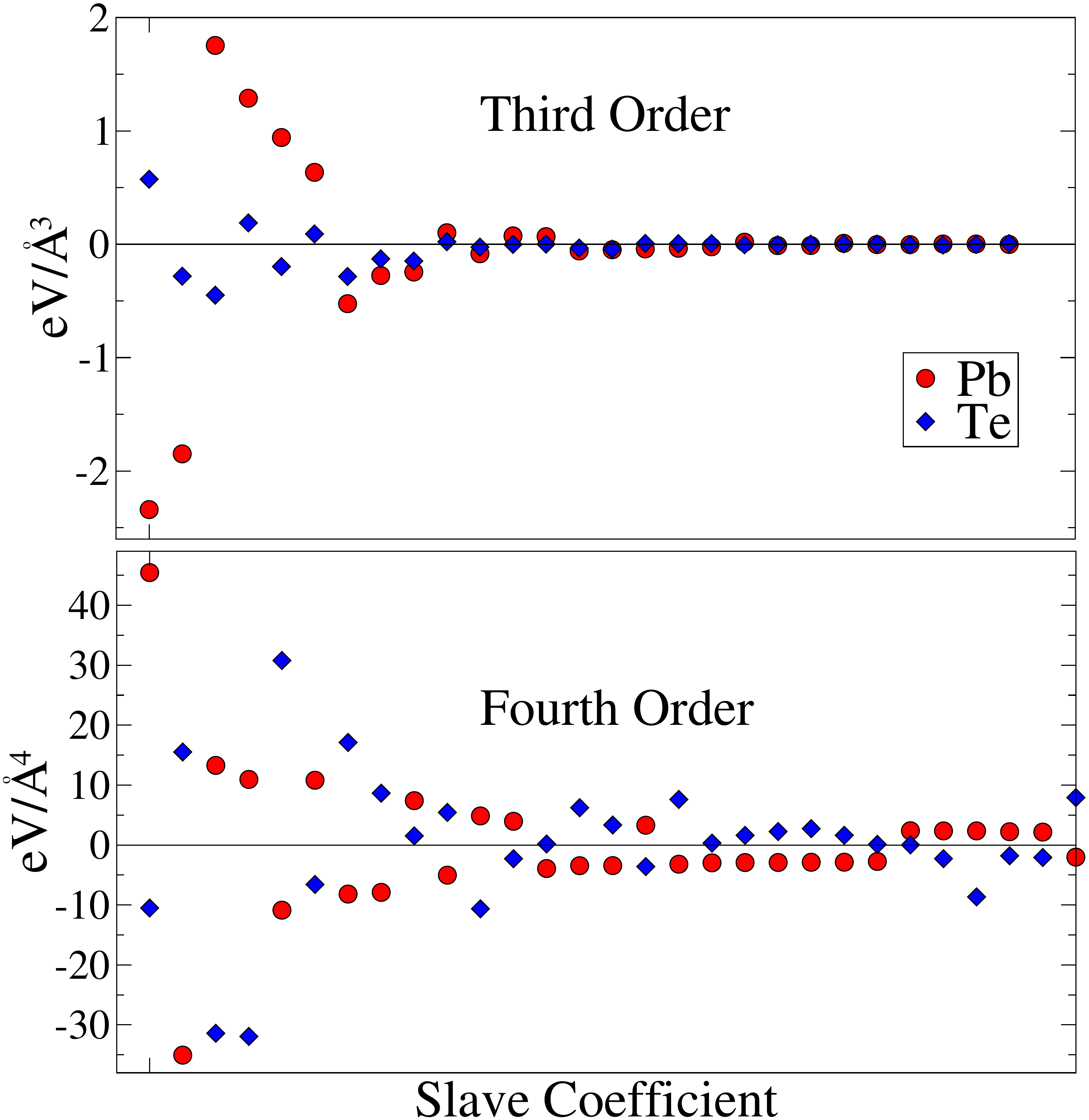}
\end{center}
\caption{A plot of the third and fourth order slave mode product coeefficients $\Phi$. The values are ordered in 
decreasing magnitude for the Pb-centered coefficients, and the same absolute ordering is used for the Te-centered coefficients.}
\label{coeff}
\end{figure}

We have computed 70 direct expansion coefficients at 3$^{rd}$ order and 427 at
4$^{th}$ order.  This exceeds the 56 slave mode coefficients at 3$^{rd}$ order
and 302 coeefficients at 4$^{th}$ order, and therefore we have an
overdetermined set of equations. Singular value decomposition can then be used
to find the optimum solution in terms of least squares, and this will yield a
unique solution for the slave mode coefficients. 
The third and fourth order terms are plotted  in figure \ref{coeff}.  At third
order, the Pb-centered slave modes have substantially larger coefficients than
the Te-centered slave modes, while the differences are less pronounced for
fourth order. The values of each slave mode coefficient are also listed in
tables \ref{term3t} and \ref{term4t}.

\section{Assessing the expansion}
Having computed the slave mode expansion coefficients up to fourth order and within next nearest neighbor interaction, 
we now evaluate the overall reliability of our expansion. The major point of concern in the method we have employed
to compute the slave mode coefficients is whether or not the slave mode expansion is sufficiently converged within the
octahedron or if non-negligible terms beyond the octahedron are present. A potent test to address this issue is to use
the slave mode expansion to compute energy, stress, and phonons as a function of lattice strain. It should be emphasized that
our slave mode expansion is performed in the absence of any strain, but if our cluster is sufficiently large the expansion will be able to 
to be used to compute the energetics under strain.  Given that strain will amplify 
the coupling to long range interactions, and that it is straightforward to compute the answer to these tests using DFT, this serves
as an ideal testbed of our method.
PbTe is sufficiently polar such that there be long range fields which will cause a  non-negligible splitting of the optical
modes near the $\Gamma$-point. These can be straightforwardly taken into account via Born effective charges\cite{Baroni2001515},
but we do not include them in this study.

\begin{figure}[htb]
\begin{center}
\includegraphics[width=\linewidth,clip= ]{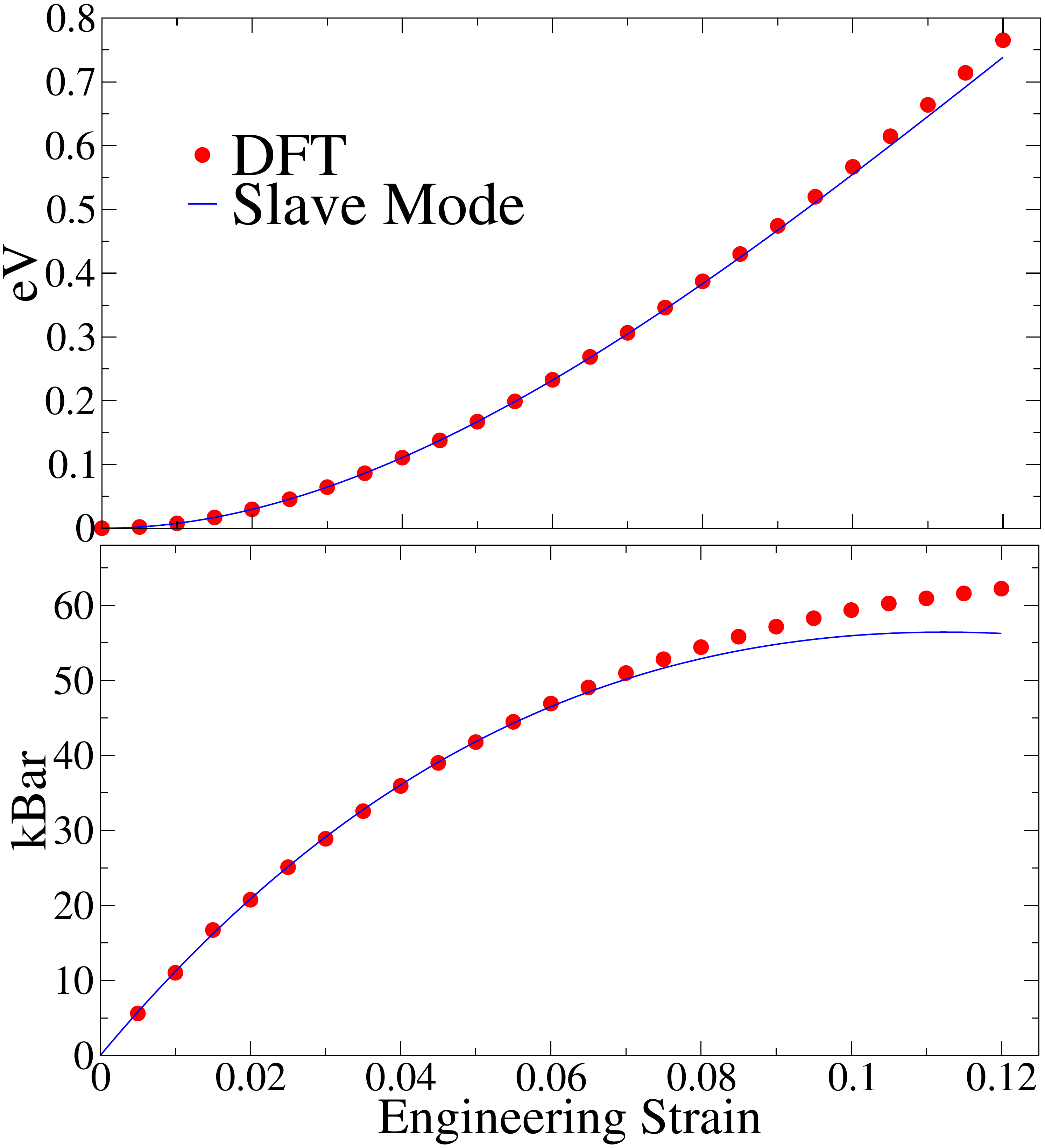}
\end{center}
\caption{Top Panel: Energy as a function of triaxial engineering strain. Bottom Panel: True Stress as a function of triaxial engineering strain.}
\label{stress}
\end{figure}

The first test is to compute the energy and the stress as a function of strain (see figure \ref{stress}). As shown,
there is remarkable agreement in the stress for strains as high as 7\% and even higher for the energy. 
At 10\% strain there is an error of roughly 8\% in the stress.  This favorable agreement suggests that longer range terms
are not substantial.

\begin{figure}[htb]
\begin{center}
\includegraphics[width=\linewidth,clip= ]{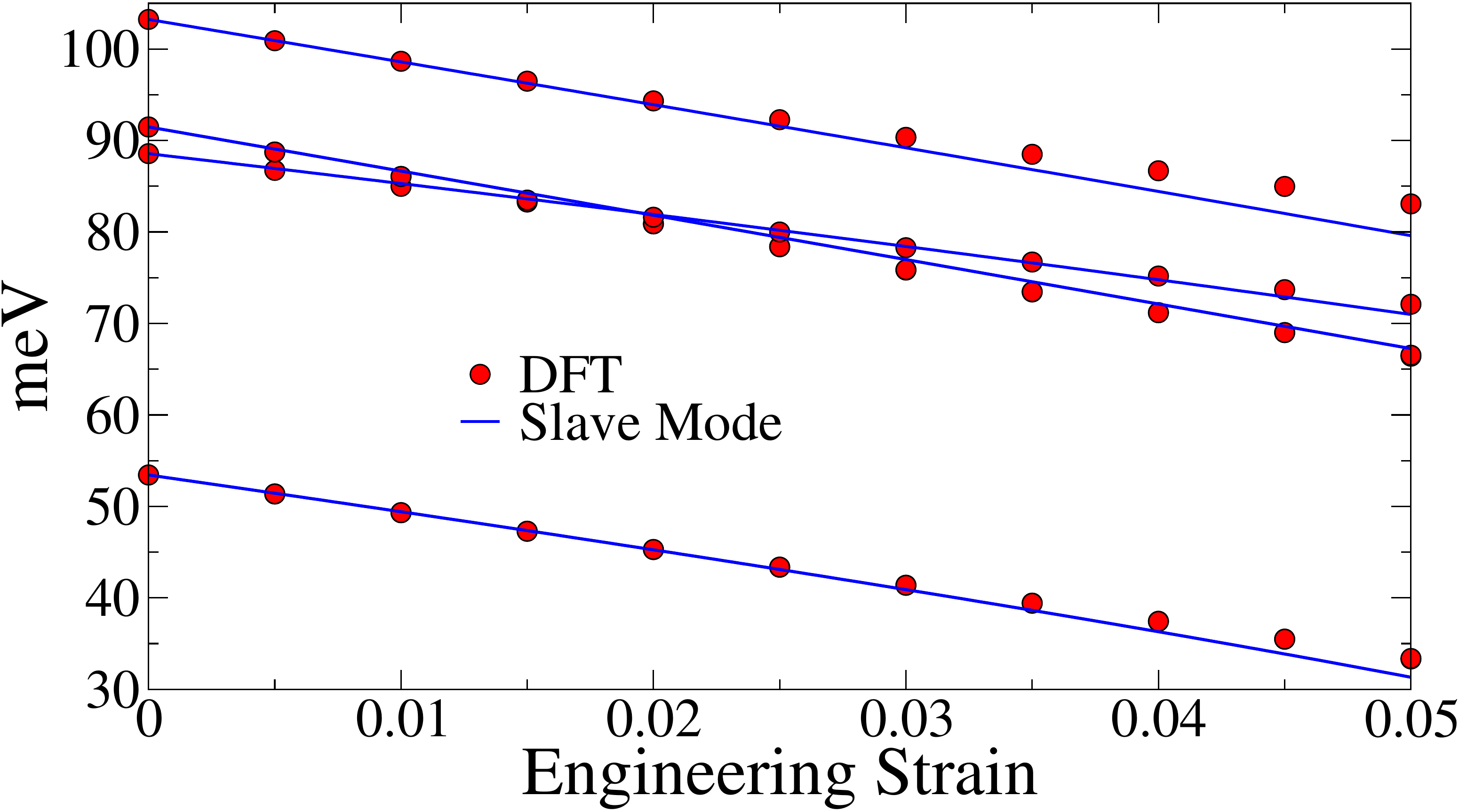}
\end{center}
\caption{L-point phonon frequencies as a function of triaxial engineering strain.}
\label{lpoint}
\end{figure}
\begin{figure}[htb]
\begin{center}
\includegraphics[width=\linewidth,clip= ]{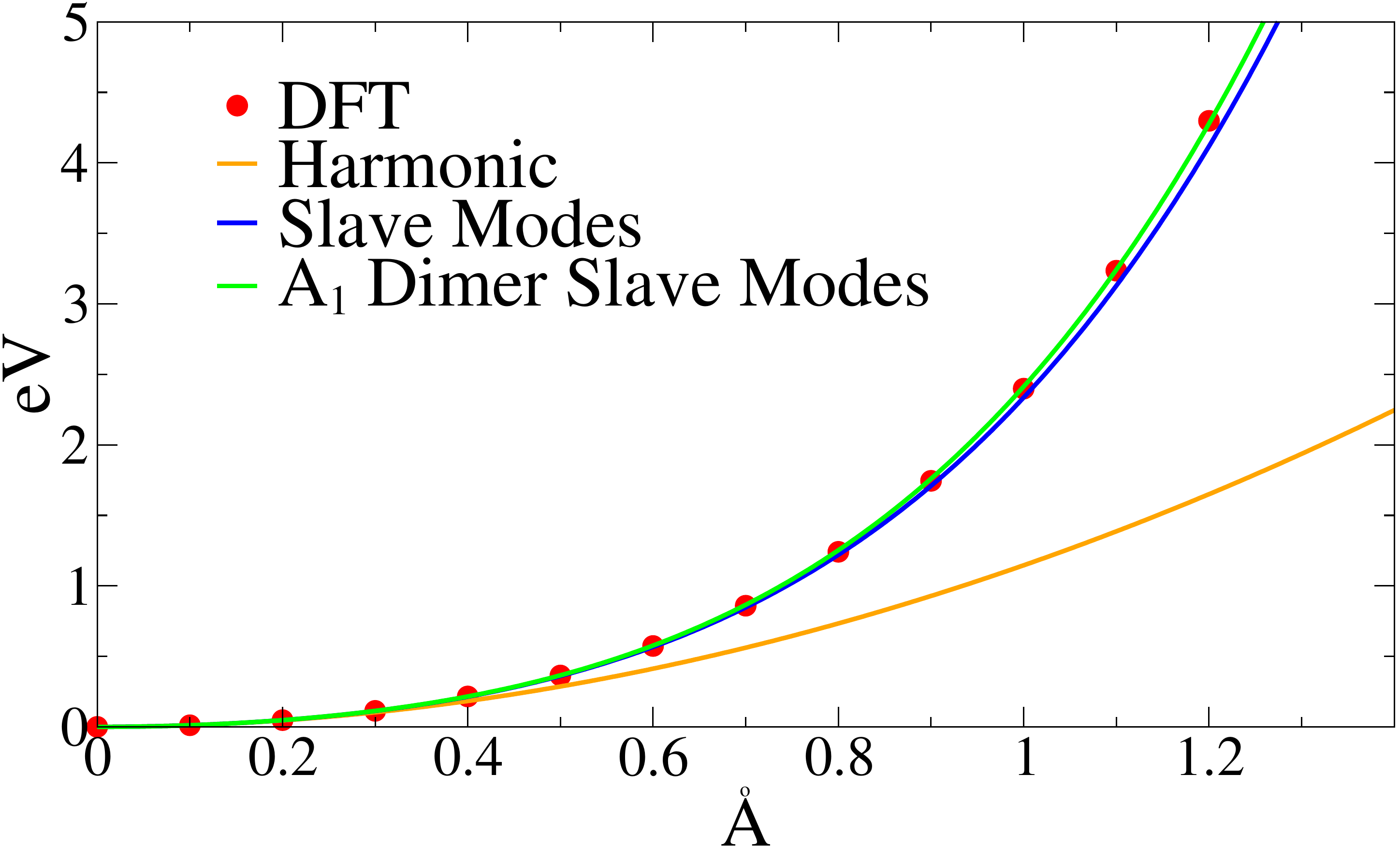}
\end{center}
\caption{Energy as a function of displacing a single Pb atom in a 216 atom supercell along the $<-3,1,1>$ direction.}
\label{singleatom}
\end{figure}

A more stringent test is to compute the phonons as a function of strain. We begin by computing the L-point phonons
as a function of strain (see figure \ref{lpoint}). As shown, there is remarable agreement up to 5\% strain.
\begin{figure}[htb]
\begin{center}
\includegraphics[width=\linewidth,clip= ]{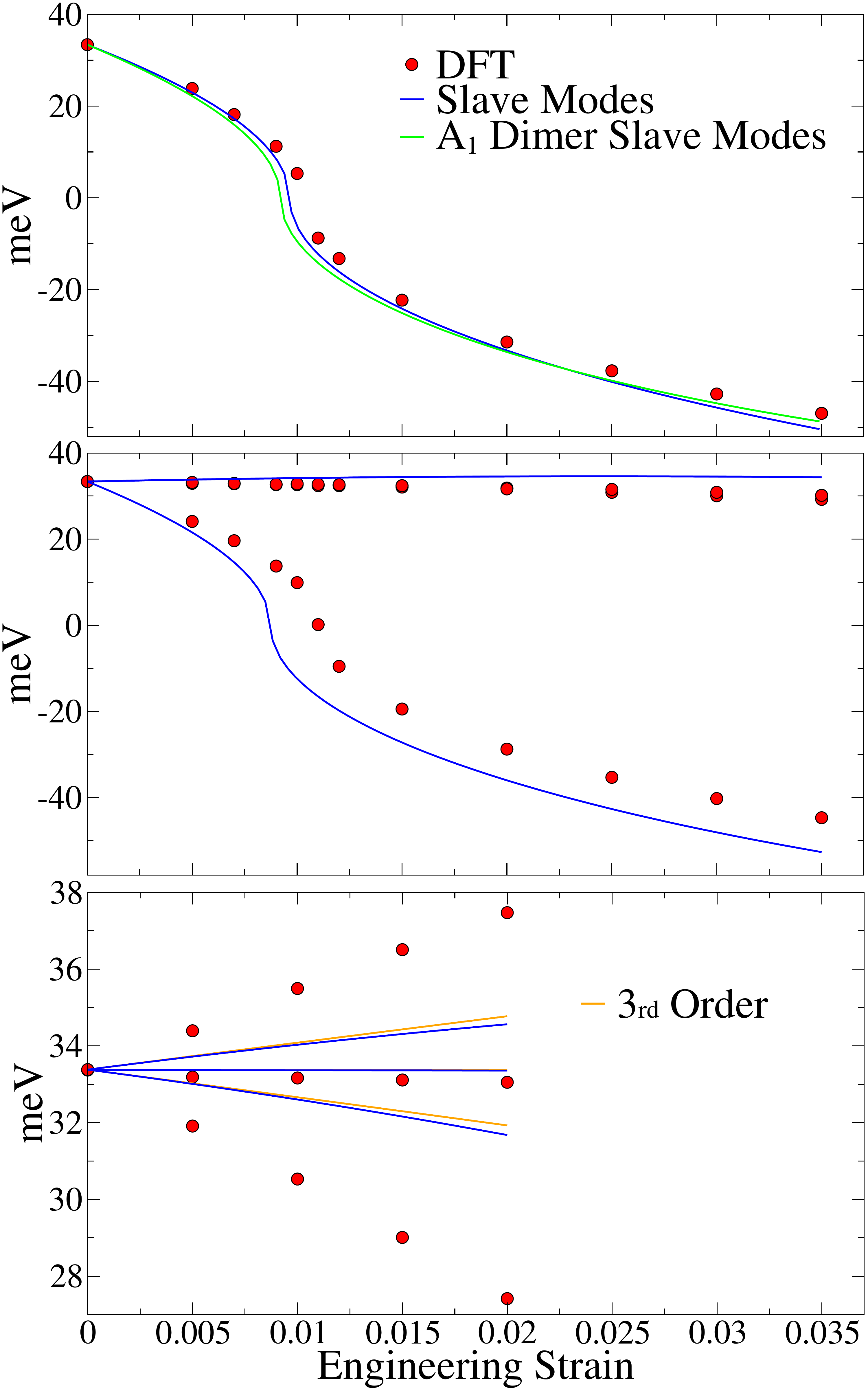}
\end{center}
\caption{$\Gamma$-point optical phonon frequencies as a function of different engineering strain states: triaxial (top panel),
uniaxial (middle panel), and shear $\gamma_{xy}$ (bottom panel).}
\label{tomode}
\end{figure}
Another test of phonons under strain is the $\Gamma$-point optical
modes. This mode is of particular interest in the context of PbTe as it
displays anomolous temperature dependence\cite{Jensen2012085313,Delaire2011614,Chen:2013}. We compute energy of
the $\Gamma$-point optical modes as a function of triaxial, uniaxial, and shear strain (see figure
\ref{tomode}). In the case of triaxial strain, the slave mode expansion
precisely captures the formation of a soft-mode. In the case of uniaxial
strain, the slave mode expansion is highly accurate for small strains and
properly captures the symmetry breaking of the optical modes. However, 
errors are apparent for the prediction of the soft mode at larger strains, though
the error is relatively constant beyond 1.5\%.  In the case of shear strain,
the splitting of the optical modes is underpredicted using the slave modes, though
the error is still within reason in this range of strain.  Nonetheless, the
troubling aspect of this result is that it does not have the correct slope in
the limit of small strains. Given that there is little difference in going from third to fourth order coefficients, this is likely a symptom of a longer range terms
that are not present in our expansion. Fortunately, the overall magnitude of this effect is rather
small, and these errors will likely be unimportant in most scenarios.
The final test will be the displacement of a single Pb atom in a 216-atom supercell (see figure \ref{singleatom}). 
The slave mode expansion is highly accurate even at displacements beyond 1.2 $\AA$. 
We believe these benchmarks demonstrate that our expansion is robust.

\section{Minimal Model}
\label{minmod}
\begin{figure}[htb]
\begin{center}
\includegraphics[width=\linewidth,clip= ]{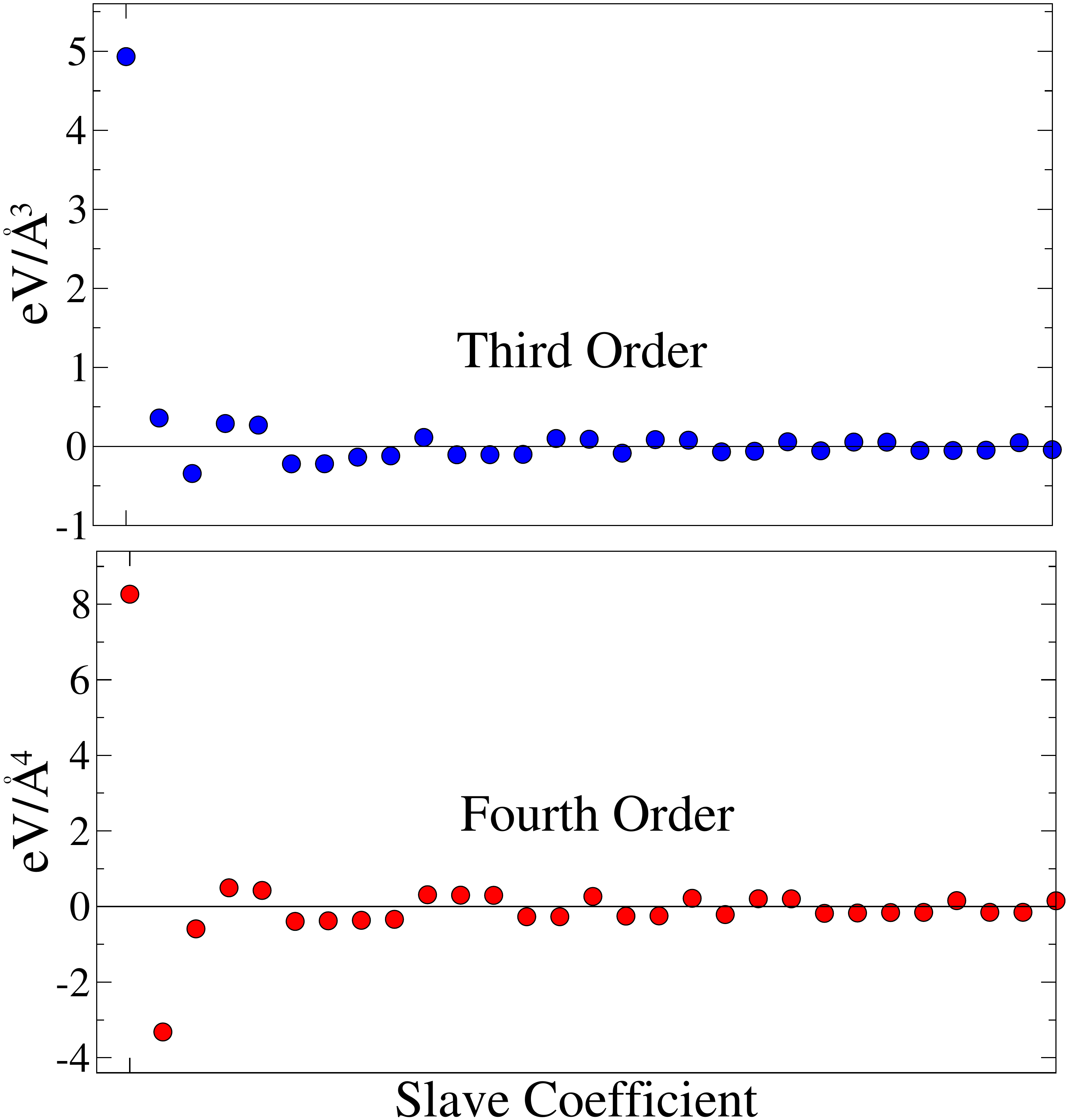}
\end{center}
\caption{A plot of the transformed third and fourth order slave mode product coeefficients $\Phi^\prime$. The values are ordered in 
decreasing magnitude.}
\label{minimal}
\end{figure}

Above we have demonstrated that our slave mode expansion accurately reproduces many key quantities. 
Nonetheless, it would be strongly desirable if we could somehow extract a \emph{minimal} model of
anharmoncity. It would be intuitive for the nearest-neighbor terms to be larger than the next nearest
neighbor terms. When choosing the octahedral cluster, the nearest and next-nearest neighbor terms will be mixed.
However, they can be seperated. 
We will
start by considering the dimer slave cluster of Pb-Te, where we will use the $C_{4v}$ symmetry along
the bond.
 Given that
this case is three dimensional, the representation for the dimer will have six
degrees of freedom, and projecting them onto the irreducible representations of
the point group yields the following representation: $\Gamma=2E\oplus 2A_1$
The representation for the modes which shift the dimer
in the $x,y,z$ directions can be chosen as one set of $ E\oplus A_{1}$ and this must be removed
leaving the following slave mode representation: $E\oplus A_{1}$. These modes
can be explicitly constructed as follows:
\begin{align}\label{dimerslaves} 
\nonumber \phi_{A_{1} }&=\frac{1}{\sqrt{2}}(u_{Te,x}-u_{Pb,x}) \\
\phi_{E^{(1)} }&=\frac{1}{\sqrt{2}}(u_{Te,y}-u_{Pb,y}) \hspace{2mm} \phi_{E^{(2)} }=\frac{1}{\sqrt{2}}(u_{Te,z}-u_{Pb,z}) 
\end{align}
In this case we chose a cluster centered on a bond where the $x$-axis aligns with the $4$-fold rotation axis.
At third order there will be two terms: $\phi_{A_{1}}^3$ and
$\phi_{A_{1}}\left(\phi_{E^{(1)}}^2+\phi_{E^{(2)}}^2   \right)$.  At
fourth order there will be four terms:$\phi_{A_{1}}^4$ and
$\phi_{A_{1}}^2\left(\phi_{E^{(1)}}^2+\phi_{E^{(2)}}^2   \right)$ and
$\phi_{E^{(1)}}^2\phi_{E^{(2)}}^2$ and
$\phi_{E^{(1)}}^4+\phi_{E^{(2)}}^4$. The $O_h$ symmetry center will then generate five more equivalent set of
slave mode products for each case, one for each bond. We can add these terms to our
original set of products in tables \ref{term3t} and \ref{term4t}, but then we
will need to remove two products at third order and four products at fourth
order to regain an irreducible space.  This is equivalent to performing a
unitary transformation within the product space. After reconstructing the
expansion coefficients for this new set of products, we then orthogonalize all
of the products to the dimer mode products. This physically motivated choice of phase convention in the 
product space achieves the goal of creating a minimal
model in that there is now one dominant term at both third and fourth order (see figure \ref{minimal}).
The dominant terms correspond to $\phi_{A_{1}}^3$ at third order and $\phi_{A_{1}}^4$ 
at fourth order.
These two terms can be used to explicitly write a minimal model for the potential (we drop for $A_1$ index below):

\begin{align}\label{potential} \nonumber
&V=V_H+ \\ \nonumber
&\Phi_{3}\sum_{\rr } 
(-\phi_{\rr x_-}^3 + \phi_{\rr x_+}^3 - \phi_{\rr y_-}^3 + \phi_{\rr y_+}^3 - \phi_{\rr z_-}^3 + \phi_{\rr z_+}^3) + \\ 
& \Phi_{4}\sum_{\rr } 
(\phi_{\rr x_-}^4 + \phi_{\rr x_+}^4 + \phi_{\rr y_-}^4 + \phi_{\rr y_+}^4 + \phi_{\rr z_-}^4 + \phi_{\rr z_+}^4)  \\ \nonumber
& \textrm{and:} \\ \nonumber
& \hspace{-2mm}\begin{array}{lr}
\phi_{\rr z_-}=\frac{1}{\sqrt{2}}(u^{\rr+\bf{a_1}}_{Te,z}      -u^\rr_{Pb,z}) &  
\phi_{\rr z_+}=\frac{1}{\sqrt{2}}(u^{\rr+\bf{a_2+a_3}}_{Te,z}  -u^\rr_{Pb,z}) \\ 
\phi_{\rr x_-}=\frac{1}{\sqrt{2}}(u^{\rr+\bf{a_2}}_{Te,x}      -u^\rr_{Pb,x}) &
\phi_{\rr x_+}=\frac{1}{\sqrt{2}}(u^{\rr+\bf{a_1+a_3}}_{Te,x}  -u^\rr_{Pb,x}) \\ 
\phi_{\rr y_-}=\frac{1}{\sqrt{2}}(u^{\rr+\bf{a_3}}_{Te,y}      -u^\rr_{Pb,y}) &
\phi_{\rr y_+}=\frac{1}{\sqrt{2}}(u^{\rr+\bf{a_1+a_2}}_{Te,y}  -u^\rr_{Pb,y}) 
\end{array}
\end{align}
Where $V_H$ is the harmonic part of the potential, $\phi$ are the slave modes 
for the dimer, $u$ are the atomic displacements, and $\bf{a_i}$
are the primitive lattice vectors of PbTe: ${\bf a_1}=a/2(1,1,0)$, ${\bf a_2}=a/2(0,1,1)$, and ${\bf a_3}=a/2(1,0,1)$.
There are six dimer slave modes per primitive unit cell, one corresponding to each Pb-Te octahedral bond,
and these are simply a displacement difference between corresponding vectors of Pb and Te.
The 
values for the expansion coefficients are found to be 
$\Phi_{3}=2.68 eV/{\AA}^{3}$ and $\Phi_{4}=3.70 eV/{\AA}^{4}$, respectively. 
We can test these two parameters by recomputing the optical modes under strain (see figure \ref{tomode}) and
the energy of displacing a single atom (see figure \ref{singleatom}), displaying excellent agreement.
This minimal model has already been used to capture the anomolous temperature dependence of the phonon spectra in PbTe\cite{Chen:2013}. 

There is one other term at fourth order which, though smaller, stands out among the other terms. This corresponds to $\phi_{A_{1}}^2\left(\phi_{E^{(1)}}^2+\phi_{E^{(2)}}^2   \right)$
and has a coefficient of $-1.37 eV/{\AA}^{4}$.

\section{Conclusions}
In conclusion, we have introduced a new approach to perform a Taylor series
expansion of the total energy as a function of the nuclear displacements.  The
novelty of our approach is the formation of new variables (ie. slave modes)
which transform like the irreducible representations of the space group while
satisfying the homogeneity of free space, and these benefits are gained at the
expense of increasing the dimensionality of the space. We used a finite
difference approach to compute the slave mode coefficients, and accurately
determined all 358 terms within fourth order and next nearest neighbor
coupling. Examining the energy, stress, and phonons under lattice strain
indicated that our expansion parameters are robust and that terms outside of
the octahedron are relatively small. Furthermore, we have introduced an
additional approach to perform a unitary transformation which allows us to
accurately compress 56 cubic terms to one term and the 302 quartic terms to one
term. This two parameter model of anharmonicity in PbTe has already been
separately used to compute the temperature dependent phonon spectrum in the
classical limit, resolving a major experimental anomaly\cite{Chen:2013}. Our slave
mode expansion should be broadly applicable to highly symmetric materials. While
substantial resources have been dedicated to characterizing minimal models of
electronic Hamiltonians, much less has been done in terms of characterizing
anharmonic interactions of relevant materials. Our approach should make this task
substantially more tractable.

\begin{acknowledgments}
CAM and XA acknowledge support from the National Science Foundation (Grant No. CMMI-1150795). YC acknowledges support from a
Columbia RISE grant.

\end{acknowledgments}

\bibliography{reference,main}

\end{document}

%% file: fig_dimer_modes.tex
\begin{tikzpicture}
% styles...
[scale=0.4,>=stealth, mycircle/.style={circle,draw=blue,fill=blue}, every node/.style={transform shape}]
% set various spacings...
\def\r{1.0};
\def\ssx{7.5};
\def\ssy{2};

% draw atoms...
\node[scale=1.4] (center) at (0,\ssy) {};
\node[mycircle] (at1)  [right=of center] {1};
\node[mycircle] (at0)  [left=of  center] {0};

% connect atoms with line...
\path[draw,dotted] (at0) -- (at1) ;
% now draw vectors...
\draw [->,thick] (at0) -- ++(\r,0) node[scale=2] at ++(-0.2,-0.5) {x$_0$};
\draw [->,thick] (at0) -- ++(0,\r) node[scale=2] at ++(0.5,-0.1) {y$_0$};
\draw [->,thick] (at1) -- ++(\r,0) node[scale=2] at ++(-0.2,-0.5) {x$_1$};
\draw [->,thick] (at1) -- ++(0,\r) node[scale=2] at ++(0.5,-0.1) {y$_1$};

\foreach \x\y\mode\moden\rax\ray\rbx\rby in {
1/0/ B$_1$           /1/   1/0/1/0,
2/0/ A$_1$           /2/  -1/0/1/0,
1/1/ B$_2$           /3/   0/1/0/1,
2/1/ A$_2$           /4/   0/-1/0/1
}{
% draw atoms...
\node[scale=1.4,label=above:\mode] (center\moden) at (\x*\ssx,\y*\ssy) {};
\node[mycircle] (at0\moden)  [ left=of center\moden] {0};
\node[mycircle] (at1\moden)  [right=of center\moden] {1};
%\node[scale=1.4,above] (label\moden) at (center\moden)  {\mode};

% connect atoms with line...
\path[draw,dotted] (at0\moden) -- (at1\moden)  ;
% now draw vectors...
\draw [->,thick] (at0\moden) -- ++(\rax*\r,\ray*\r);
\draw [->,thick] (at1\moden) -- ++(\rbx*\r,\rby*\r);
}

% loop is done... put a red X through the ones that we will throw out...
%\begin{pgfonlayer}{background}
\node[circle,scale=4,red] at (center1) {X};
\node[circle,scale=4,red] at (center3) {X};
%\end{pgfonlayer}

\end{tikzpicture}

%% file: fig_square_modes.tex
\begin{tikzpicture}
% styles...
[scale=0.4,>=stealth, mycircle/.style={circle,draw=blue,fill=blue}, every node/.style={transform shape}]
% set various spacings...
\def\r{2.0};
\def\ssx{7.5};
\def\ssy{6};
% perform a loop over all the modes...
\foreach \x\y\mode\moden\rax\ray\rbx\rby\rcx\rcy\rdx\rdy in {
1/2/ A$_1$           /1/   0.50 /  0.50 /  0.50 / -0.50 / -0.50 / -0.50 / -0.50 /  0.50,
0/1/ B$_1$           /2/   0.50 / -0.50 /  0.50 /  0.50 / -0.50 /  0.50 / -0.50 / -0.50,
2/2/ A$_2$           /3/   0.50 / -0.50 / -0.50 / -0.50 / -0.50 /  0.50 /  0.50 /  0.50,
0/0/ B$_2$           /4/   0.50 /  0.50 / -0.50 /  0.50 / -0.50 / -0.50 /  0.50 / -0.50,
1/1/ E$^{(1)}$       /5/   0.50 /  0.50 / -0.50 / -0.50 /  0.50 /  0.50 / -0.50 / -0.50,
2/1/ E$^{(2)}$       /6/   0.50 / -0.50 / -0.50 /  0.50 /  0.50 / -0.50 / -0.50 /  0.50,
1/0/ E$^{\prime(1)}$ /7/   0.50 /  0.50 /  0.50 /  0.50 /  0.50 /  0.50 /  0.50 /  0.50,
2/0/ E$^{\prime(2)}$ /8/  -0.50 /  0.50 / -0.50 /  0.50 / -0.50 /  0.50 / -0.50 /  0.50
}{
% draw atoms...
\node[scale=1.4] (center\moden) at (\x*\ssx,\y*\ssy) {\mode};
\node[mycircle] (at0\moden)  [above right=of center\moden] {0};
\node[mycircle] (at1\moden)  [below right=of center\moden] {1};
\node[mycircle] (at2\moden)  [below left=of  center\moden] {2};
\node[mycircle] (at3\moden)  [above left=of  center\moden] {3};

% connect atoms with line...
\path[draw,dotted] (at0\moden) -- (at1\moden) -- (at2\moden) -- (at3\moden) -- (at0\moden) ;
% now draw vectors...
\draw [->,thick] (at0\moden) -- ++(\rax*\r,\ray*\r);
\draw [->,thick] (at1\moden) -- ++(\rbx*\r,\rby*\r);
\draw [->,thick] (at2\moden) -- ++(\rcx*\r,\rcy*\r);
\draw [->,thick] (at3\moden) -- ++(\rdx*\r,\rdy*\r);
}
% loop is done... put a red X through the ones that we will throw out...
\node[circle,scale=4,red] at (center7) {X};
\node[circle,scale=4,red] at (center8) {X};

% draw the reference square...
\node[scale=1.4] (center) at (0,2*\ssy) {};
\node[mycircle] (at0)  [above right=of center] {0};
\node[mycircle] (at1)  [below right=of center] {1};
\node[mycircle] (at2)  [below left=of  center] {2};
\node[mycircle] (at3)  [above left=of  center] {3};

% connect atoms with line...
\path[draw,dotted] (at0) -- (at1) -- (at2) -- (at3) -- (at0) ;
% now draw vectors...
\def\r{1.0};
\draw [->,thick] (at0) -- ++(\r,0) node[scale=2.5] at ++(0.2,-0.45) {x$_0$};
\draw [->,thick] (at0) -- ++(0,\r) node[scale=2.5] at ++(0.6,-0.1 ) {y$_0$};
\draw [->,thick] (at1) -- ++(\r,0) node[scale=2.5] at ++(0.2,-0.45) {x$_1$};
\draw [->,thick] (at1) -- ++(0,\r) node[scale=2.5] at ++(0.6,-0.1 ) {y$_1$};
\draw [->,thick] (at2) -- ++(\r,0) node[scale=2.5] at ++(0.2,-0.45) {x$_2$};
\draw [->,thick] (at2) -- ++(0,\r) node[scale=2.5] at ++(0.6,-0.1 ) {y$_2$};
\draw [->,thick] (at3) -- ++(\r,0) node[scale=2.5] at ++(0.2,-0.45) {x$_3$};
\draw [->,thick] (at3) -- ++(0,\r) node[scale=2.5] at ++(0.6,-0.1 ) {y$_3$};
\end{tikzpicture}

%% file: square_lattice_overlap.tex
\begin{tikzpicture}
% styles...
[scale=0.4,>=stealth, mycircle/.style={circle,draw=blue,fill=blue}, every node/.style={transform shape}]
% set various spacings...
\def\ssx{2.4};
\def\ssy{2.4};

\foreach \x/\y/\mode in {
0/0/,1/0/,2/0/,3/0/,4/0/,5/0/,0/1/,1/1/,2/1/,3/1/,4/1/,5/1/,0/2/,1/2/,2/2,3/2/,4/2/,5/2/,0/3/,1/3/,2/3/,
3/3/,4/3/,5/3/,0/4/,1/4/,2/4/,3/4/,4/4/,5/4/,0/5/,1/5/,2/5/,3/5/,4/5/,5/5/
}{
% draw atoms...
\node[scale=1.4,mycircle] (at\x\y) at (\x*\ssx,\y*\ssy) {};
}
% index atoms
\node[scale=1.7, above right] at (7.4,7.4)  {0};
\node[scale=1.7, below right] at (7.4,4.6)  {1};
\node[scale=1.7, below left] at (4.6,4.6)  {2};
\node[scale=1.7, above left] at (4.6,7.4)  {3};
% circle out atom sets
\draw [cyan,thick] (6,6) circle [radius=2.2];
\draw [lime,thick] (6,8.4) circle [radius=2.2];
\draw [lime,thick] (6,3.6) circle [radius=2.2];
\draw [lime,thick] (8.4,6) circle [radius=2.2];
\draw [lime,thick] (3.6,6) circle [radius=2.2];
\draw [yellow,thick] (8.4,8.4) circle [radius=2.2];
\draw [yellow,thick] (3.6,8.4) circle [radius=2.2];
\draw [yellow,thick] (8.4,3.6) circle [radius=2.2];
\draw [yellow,thick] (3.6,3.6) circle [radius=2.2];
\end{tikzpicture}

%% file: table_coeff3.tex
\\ \hline$T_{2g} \otimes 2T_{1u} \otimes T_{2u}$ & 32, 45 & -0.059,-0.048 & -0.033,-0.043 
\\ \hline$E_g \otimes T_{1g} \otimes T_{1g}$ & 1 & -0.011 & 0.001 
\\ \hline$T_{1g} \otimes T_{1g} \otimes T_{2g}$ & 16 & 0.002 & -0.002 
\\ \hline$A_{1g} \otimes E_g \otimes E_g$ & 1 & 0.074 & -0.002 
\\ \hline$T_{2g} \otimes 2T_{1u} \otimes 2T_{1u}$ & 38, 59, 84 & -0.276,-0.245,-0.524 & -0.129,-0.148,-0.284 
\\ \hline$T_{1g} \otimes 2T_{1u} \otimes T_{2u}$ & 45, 54 & 0.102,-0.084 & 0.022,-0.025 
\\ \hline$A_{1g} \otimes T_{2u} \otimes T_{2u}$ & 5 & -0.01 & N/A
\\ \hline$T_{2g} \otimes T_{2g} \otimes T_{2g}$ & 6 & -0.003 & 0.002 
\\ \hline$E_g \otimes 2T_{1u} \otimes T_{2u}$ & 20, 29 & -0.041,0.067 & 0.008,0.001 
\\ \hline$A_{1g} \otimes 2T_{1u} \otimes 2T_{1u}$ & 4, 15, 22 & -1.849,1.288,0.635 & -0.282,0.188,0.091 
\\ \hline$E_g \otimes T_{2g} \otimes T_{2g}$ & 14 & -0.003 & -0.006 
\\ \hline$T_{1g} \otimes 2T_{1u} \otimes 2T_{1u}$ & 6 & -0.022 & 0.006 
\\ \hline$E_g \otimes E_g \otimes E_g$ & 4 & -0.035 & 0.005 
\\ \hline$E_g \otimes 2T_{1u} \otimes 2T_{1u}$ & 4, 22, 51 & -2.341,0.941,1.754 & 0.573,-0.197,-0.449 
\\ \hline$T_{2g} \otimes T_{2u} \otimes T_{2u}$ & 2 & 0.002 & -0.007 
\\ \hline$E_g \otimes T_{1g} \otimes T_{2g}$ & 5 & 0.018 & -0.005 
\\ \hline$A_{1g} \otimes T_{2g} \otimes T_{2g}$ & 9 & -0.002 & 0.006 
\\ \hline$A_{1g} \otimes A_{1g} \otimes A_{1g}$ & 1 & 0.01 & N/A
\\ \hline$A_{1g} \otimes T_{1g} \otimes T_{1g}$ & 1 & -0.011 & -0.004 
\\ \hline$E_g \otimes T_{2u} \otimes T_{2u}$ & 10 & 0.008 & 0.003

%% file: table_coeff4_withseeds.tex
$E_{g} \otimes E_{g} \otimes T_{2g} \otimes T_{2g} $ & 14, 32 & -0.043, -0.034 & 0.008, 0.01\\ \hline
$T_{2g} \otimes T_{2g} \otimes 2T_{1u} \otimes T_{2u} $ & 83, 151, 36, 27 & 0.938, -0.681, 0.031, -0.05 & -1.014, 0.742, 0.088, -0.077\\ \hline
$T_{2u} \otimes T_{2u} \otimes T_{2u} \otimes T_{2u} $ & 1, 5 & 0.017, 0.003 & -0.015, 0.001\\ \hline
$E_{g} \otimes T_{1g} \otimes T_{1g} \otimes T_{2g} $ & 35 & -0.0 & 0.003\\ \hline
$T_{1g} \otimes T_{1g} \otimes T_{1g} \otimes T_{2g} $ & 25 & -0.007 & -0.001\\ \hline
$A_{1g} \otimes T_{2g} \otimes T_{2g} \otimes T_{2g} $ & 6 & 0.001 & 0.002\\ \hline
$A_{1g} \otimes E_{g} \otimes T_{1g} \otimes T_{1g} $ & 14 & 0.029 & -0.01\\ \hline
$T_{1g} \otimes T_{2g} \otimes T_{2g} \otimes T_{2g} $ & 9 & -0.002 & 0.007\\ \hline
$E_{g} \otimes E_{g} \otimes T_{1g} \otimes T_{2g} $ & 14 & -0.121 & 0.047\\ \hline
$2T_{1u}  \otimes 2T_{1u}  \otimes 2T_{1u}  \otimes 2T_{1u} $ & 533, 1, 1044, 11, 22, 59, 606, 130, 15, 1037, 522  & -8.18, 7.405, -3.203, 10.942, 45.45, -10.874, 13.284, -35.053, -2.019, 10.804, -28.53 & 17.113, 1.528, 7.597, -31.941, -10.505, 30.786, -31.418, 15.514, 7.908, -6.578,  N/A \\ \hline
$T_{1g} \otimes T_{1g} \otimes T_{2g} \otimes T_{2g} $ & 73, 41, 11 & 0.003, 0.102, -0.005 & 0.0, -0.102, 0.003\\ \hline
$A_{1g} \otimes A_{1g} \otimes A_{1g} \otimes A_{1g} $ & 1 & 0.002 & -0.001\\ \hline
$E_{g} \otimes T_{1g} \otimes T_{2g} \otimes T_{2g} $ & 47 & 0.002 & 0.005\\ \hline
$T_{2g} \otimes T_{2g} \otimes T_{2g} \otimes T_{2g} $ & 41, 5 & 0.017, 0.001 & -0.017, -0.002\\ \hline
$T_{1g} \otimes T_{1g} \otimes T_{1g} \otimes T_{1g} $ & 41, 45 & 0.018, 0.003 & -0.014, 0.001\\ \hline
$E_{g} \otimes T_{1g} \otimes 2T_{1u} \otimes T_{2u} $ & 54, 108, 1, 8 & 0.172, -0.297, 0.467, -0.001 & -0.11, 0.195, -0.304, -0.006\\ \hline
$E_{g} \otimes T_{2g} \otimes 2T_{1u} \otimes 2T_{1u} $ & 89, 16, 146, 102 & -1.391, 1.118, -0.416, 0.398 & 1.716, -0.56, 0.801, -0.685\\ \hline
$2T_{1u} \otimes T_{2u} \otimes T_{2u} \otimes T_{2u} $ & 18, 83 & 0.038, 0.054 & -0.018, -0.018\\ \hline
$T_{1g} \otimes T_{1g} \otimes 2T_{1u} \otimes 2T_{1u} $ & 317, 53, 29, 95, 292, 152, 183, 155, 289 & 2.214, 0.136, 0.021, 0.089, -0.004, 0.833, 0.051, -2.914, -0.013 & -1.801, -0.238, -0.05, -0.137, 0.092, -0.584, -0.107, 2.242, -0.046\\ \hline
$A_{1g} \otimes T_{2g} \otimes T_{2u} \otimes T_{2u} $ & 21 & 0.007 & 0.005\\ \hline
$E_{g} \otimes E_{g} \otimes E_{g} \otimes E_{g} $ & 4 & 0.024 & -0.01\\ \hline
$E_{g} \otimes T_{1g} \otimes 2T_{1u} \otimes 2T_{1u} $ & 16, 192, 111, 167 & 1.062, 3.955, -2.932, -2.859 & -0.674, -2.289, 1.606, 1.627\\ \hline
$A_{1g} \otimes T_{1g} \otimes T_{1g} \otimes T_{2g} $ & 8 & -0.001 & 0.002\\ \hline
$T_{1g} \otimes T_{1g} \otimes T_{2u} \otimes T_{2u} $ & 77, 5, 12 & 0.006, 0.102, -0.005 & 0.001, -0.098, -0.002\\ \hline
$E_{g} \otimes T_{2g} \otimes 2T_{1u} \otimes T_{2u} $ & 17, 99, 71, 21 & 0.226, 0.207, -0.204, 0.248 & -0.069, 0.098, 0.198, -0.363\\ \hline
$E_{g} \otimes E_{g} \otimes T_{1g} \otimes T_{1g} $ & 9, 28 & 0.071, -0.106 & -0.026, 0.039\\ \hline
$A_{1g} \otimes T_{1g} \otimes 2T_{1u} \otimes T_{2u} $ & 10, 40 & -0.263, -0.359 & 0.071, 0.118\\ \hline
$A_{1g} \otimes T_{2g} \otimes 2T_{1u} \otimes 2T_{1u} $ & 102, 38, 41 & -0.905, 0.882, -1.902 & 0.675, -0.724, 1.459\\ \hline
$T_{1g} \otimes T_{1g} \otimes 2T_{1u} \otimes T_{2u} $ & 83, 39, 107, 74 & 0.986, -0.009, 0.025, -0.718 & -0.918, 0.02, -0.027, 0.662\\ \hline
$E_{g} \otimes T_{2g} \otimes T_{2u} \otimes T_{2u} $ & 29 & -0.003 & 0.002\\ \hline
$2T_{1u} \otimes 2T_{1u} \otimes 2T_{1u} \otimes T_{2u} $ & 30, 540, 280, 306, 142, 7 & 2.355, -1.842, -0.811, 4.857, -3.446, -1.88 & -8.661, 4.674, 3.853, -10.649, 6.226, 4.979\\ \hline
$A_{1g} \otimes A_{1g} \otimes 2T_{1u} \otimes 2T_{1u} $ & 8, 11, 29 & 0.598, -0.725, 0.145 & 0.078, -0.158, 0.096\\ \hline
$A_{1g} \otimes E_{g} \otimes E_{g} \otimes E_{g} $ & 1 & 0.03 & 0.0\\ \hline
$A_{1g} \otimes E_{g} \otimes T_{2u} \otimes T_{2u} $ & 5 & 0.028 & -0.005\\ \hline
$A_{1g} \otimes E_{g} \otimes T_{1g} \otimes T_{2g} $ & 10 & 0.17 & -0.06\\ \hline
$A_{1g} \otimes E_{g} \otimes 2T_{1u} \otimes 2T_{1u} $ & 29, 54, 51 & -2.974, -3.915, 2.377 & 0.315, 0.165, 0.021\\ \hline
$A_{1g} \otimes A_{1g} \otimes T_{1g} \otimes T_{1g} $ & 5 & -0.018 & 0.001\\ \hline
$A_{1g} \otimes T_{1g} \otimes 2T_{1u} \otimes 2T_{1u} $ & 89 & 0.831 & -0.347\\ \hline
$E_{g}  \otimes E_{g}  \otimes 2T_{1u}  \otimes 2T_{1u} $ & 130, 47, 72, 11, 15 116  & -0.093, -2.737, -0.91, -0.678, -0.85, 1.64 & 0.255, 0.112, -0.052, -0.397, 0.175,  N/A \\ \hline
$A_{1g} \otimes A_{1g} \otimes T_{2g} \otimes T_{2g} $ & 1 & -0.013& N/A\\ \hline
$A_{1g} \otimes T_{2g} \otimes 2T_{1u} \otimes T_{2u} $ & 10, 23 & -0.146, -0.237 & 0.108, 0.16\\ \hline
$E_{g} \otimes E_{g} \otimes T_{2u} \otimes T_{2u} $ & 18, 36 & 0.043, -0.037 & -0.009, 0.007\\ \hline
$A_{1g} \otimes E_{g} \otimes 2T_{1u} \otimes T_{2u} $ & 34, 20 & 0.754, -0.283 & -0.344, 0.133\\ \hline
$T_{1g} \otimes T_{2g} \otimes 2T_{1u} \otimes T_{2u} $ & 83, 74, 22, 143, 155, 72, 79, 113 & 1.913, -1.372, -0.02, 0.04, -0.07, -0.004, -0.056, -0.026 & -2.0, 1.461, -0.015, -0.052, 0.039, -0.027, 0.029, 0.038\\ \hline
$E_{g} \otimes T_{1g} \otimes T_{2u} \otimes T_{2u} $ & 2 & -0.001 & -0.001\\ \hline
$T_{2g} \otimes T_{2g} \otimes T_{2u} \otimes T_{2u} $ & 24, 77, 41 & -0.001, 0.003, 0.098 & -0.007, -0.002, -0.105\\ \hline
$2T_{1u} \otimes 2T_{1u} \otimes T_{2u} \otimes T_{2u} $ & 198, 99, 28, 264, 135, 72, 87, 20, 261 & -0.194, -3.438, 0.63, 0.415, -0.47, 1.251, 0.813, 0.409, 2.36 & 0.326, 3.335, -0.834, 0.041, 0.531, -1.225, 0.133, 0.09, -2.28\\ \hline
$T_{2g}  \otimes T_{2g}  \otimes 2T_{1u}  \otimes 2T_{1u} $ & 306, 196, 1, 317, 66, 8, 310, 162, 45  & -2.88, -0.216, 0.857, 2.154, -0.093, -0.048, -0.015, 0.067, -0.118 & 2.73, -0.047, -0.8, -2.083, -0.056, -0.106, -0.125, 0.226, N/A \\ \hline
$T_{1g} \otimes T_{2g} \otimes 2T_{1u} \otimes 2T_{1u} $ & 15, 95, 120, 221, 18, 324, 117 & 3.295, -0.036, -0.071, -0.081, -7.9, -5.024, 0.104 & -3.6, -0.228, -0.166, -0.148, 8.646, 5.441, 0.098\\ \hline
$A_{1g} \otimes A_{1g} \otimes T_{2u} \otimes T_{2u} $ & 1 & -0.014 & -0.0\\ \hline
$A_{1g} \otimes E_{g} \otimes T_{2g} \otimes T_{2g} $ & 9 & 0.018 & -0.013\\ \hline
$T_{1g} \otimes T_{2g} \otimes T_{2u} \otimes T_{2u} $ & 60, 73 & 0.002, -0.202 & -0.005, 0.205\\ \hline
$A_{1g} \otimes A_{1g} \otimes E_{g} \otimes E_{g} $ & 4 & 0.027 & 0.003\\ \hline
$E_{g} \otimes E_{g} \otimes 2T_{1u} \otimes T_{2u} $ & 6, 29 & -0.288, -0.326 & -0.008, 0.021

%% file: potential.bbl
\begin{thebibliography}{43}%
\makeatletter
\providecommand \@ifxundefined [1]{%
 \@ifx{#1\undefined}
}%
\providecommand \@ifnum [1]{%
 \ifnum #1\expandafter \@firstoftwo
 \else \expandafter \@secondoftwo
 \fi
}%
\providecommand \@ifx [1]{%
 \ifx #1\expandafter \@firstoftwo
 \else \expandafter \@secondoftwo
 \fi
}%
\providecommand \natexlab [1]{#1}%
\providecommand \enquote  [1]{``#1''}%
\providecommand \bibnamefont  [1]{#1}%
\providecommand \bibfnamefont [1]{#1}%
\providecommand \citenamefont [1]{#1}%
\providecommand \href@noop [0]{\@secondoftwo}%
\providecommand \href [0]{\begingroup \@sanitize@url \@href}%
\providecommand \@href[1]{\@@startlink{#1}\@@href}%
\providecommand \@@href[1]{\endgroup#1\@@endlink}%
\providecommand \@sanitize@url [0]{\catcode `\\12\catcode `\$12\catcode
  `\&12\catcode `\#12\catcode `\^12\catcode `\_12\catcode `\%12\relax}%
\providecommand \@@startlink[1]{}%
\providecommand \@@endlink[0]{}%
\providecommand \url  [0]{\begingroup\@sanitize@url \@url }%
\providecommand \@url [1]{\endgroup\@href {#1}{\urlprefix }}%
\providecommand \urlprefix  [0]{URL }%
\providecommand \Eprint [0]{\href }%
\providecommand \doibase [0]{http://dx.doi.org/}%
\providecommand \selectlanguage [0]{\@gobble}%
\providecommand \bibinfo  [0]{\@secondoftwo}%
\providecommand \bibfield  [0]{\@secondoftwo}%
\providecommand \translation [1]{[#1]}%
\providecommand \BibitemOpen [0]{}%
\providecommand \bibitemStop [0]{}%
\providecommand \bibitemNoStop [0]{.\EOS\space}%
\providecommand \EOS [0]{\spacefactor3000\relax}%
\providecommand \BibitemShut  [1]{\csname bibitem#1\endcsname}%
\let\auto@bib@innerbib\@empty
%</preamble>
\bibitem [{\citenamefont {Baroni}\ \emph {et~al.}(2001)\citenamefont {Baroni},
  \citenamefont {deGironcoli}, \citenamefont {DalCorso},\ and\ \citenamefont
  {Giannozzi}}]{Baroni2001515}%
  \BibitemOpen
  \bibfield  {author} {\bibinfo {author} {\bibfnamefont {S.}~\bibnamefont
  {Baroni}}, \bibinfo {author} {\bibfnamefont {S.}~\bibnamefont {deGironcoli}},
  \bibinfo {author} {\bibfnamefont {A.}~\bibnamefont {DalCorso}}, \ and\
  \bibinfo {author} {\bibfnamefont {P.}~\bibnamefont {Giannozzi}},\ }\href@noop
  {} {\bibfield  {journal} {\bibinfo  {journal} {Rev. Mod. Phys.}\ }\textbf
  {\bibinfo {volume} {73}},\ \bibinfo {pages} {515} (\bibinfo {year}
  {2001})}\BibitemShut {NoStop}%
\bibitem [{\citenamefont {Kunc}\ and\ \citenamefont
  {Martin}(1982)}]{Kunc1982406}%
  \BibitemOpen
  \bibfield  {author} {\bibinfo {author} {\bibfnamefont {K.}~\bibnamefont
  {Kunc}}\ and\ \bibinfo {author} {\bibfnamefont {R.~M.}\ \bibnamefont
  {Martin}},\ }\href {\doibase 10.1103/PhysRevLett.48.406} {\bibfield
  {journal} {\bibinfo  {journal} {Phys. Rev. Lett.}\ }\textbf {\bibinfo
  {volume} {48}},\ \bibinfo {pages} {406} (\bibinfo {year} {1982})}\BibitemShut
  {NoStop}%
\bibitem [{\citenamefont {Alfe}(2009)}]{Alfe20092622}%
  \BibitemOpen
  \bibfield  {author} {\bibinfo {author} {\bibfnamefont {D.}~\bibnamefont
  {Alfe}},\ }\href@noop {} {\bibfield  {journal} {\bibinfo  {journal} {Computer
  Physics Communications}\ }\textbf {\bibinfo {volume} {180}},\ \bibinfo
  {pages} {2622} (\bibinfo {year} {2009})}\BibitemShut {NoStop}%
\bibitem [{\citenamefont {Parlinski}\ \emph {et~al.}(1997)\citenamefont
  {Parlinski}, \citenamefont {Li},\ and\ \citenamefont
  {Kawazoe}}]{Parlinski19974063}%
  \BibitemOpen
  \bibfield  {author} {\bibinfo {author} {\bibfnamefont {K.}~\bibnamefont
  {Parlinski}}, \bibinfo {author} {\bibfnamefont {Z.}~\bibnamefont {Li}}, \
  and\ \bibinfo {author} {\bibfnamefont {Y.}~\bibnamefont {Kawazoe}},\
  }\href@noop {} {\bibfield  {journal} {\bibinfo  {journal} {Phys. Rev. Lett.}\
  }\textbf {\bibinfo {volume} {78}},\ \bibinfo {pages} {4063} (\bibinfo {year}
  {1997})}\BibitemShut {NoStop}%
\bibitem [{\citenamefont {VandeWalle}\ and\ \citenamefont
  {Ceder}(2002)}]{Walle200211}%
  \BibitemOpen
  \bibfield  {author} {\bibinfo {author} {\bibfnamefont {A.}~\bibnamefont
  {VandeWalle}}\ and\ \bibinfo {author} {\bibfnamefont {G.}~\bibnamefont
  {Ceder}},\ }\href@noop {} {\bibfield  {journal} {\bibinfo  {journal} {Rev.
  Mod. Phys.}\ }\textbf {\bibinfo {volume} {74}},\ \bibinfo {pages} {11}
  (\bibinfo {year} {2002})}\BibitemShut {NoStop}%
\bibitem [{\citenamefont {Papaconstantopoulos}\ and\ \citenamefont
  {Mehl}(2003)}]{Papaconstantopoulos2003413}%
  \BibitemOpen
  \bibfield  {author} {\bibinfo {author} {\bibfnamefont {D.~A.}\ \bibnamefont
  {Papaconstantopoulos}}\ and\ \bibinfo {author} {\bibfnamefont {M.~J.}\
  \bibnamefont {Mehl}},\ }\href@noop {} {\bibfield  {journal} {\bibinfo
  {journal} {Journal Of Physics-condensed Matter}\ }\textbf {\bibinfo {volume}
  {15}},\ \bibinfo {pages} {R413} (\bibinfo {year} {2003})}\BibitemShut
  {NoStop}%
\bibitem [{\citenamefont {Bowler}\ and\ \citenamefont
  {Miyazaki}(2012)}]{Bowler2012036503}%
  \BibitemOpen
  \bibfield  {author} {\bibinfo {author} {\bibfnamefont {D.~R.}\ \bibnamefont
  {Bowler}}\ and\ \bibinfo {author} {\bibfnamefont {T.}~\bibnamefont
  {Miyazaki}},\ }\href@noop {} {\bibfield  {journal} {\bibinfo  {journal}
  {Reports On Progress In Physics}\ }\textbf {\bibinfo {volume} {75}},\
  \bibinfo {pages} {036503} (\bibinfo {year} {2012})}\BibitemShut {NoStop}%
\bibitem [{\citenamefont {Vandevondele}\ \emph {et~al.}(2012)\citenamefont
  {Vandevondele}, \citenamefont {Borstnik},\ and\ \citenamefont
  {Hutter}}]{Vandevondele20123565}%
  \BibitemOpen
  \bibfield  {author} {\bibinfo {author} {\bibfnamefont {J.}~\bibnamefont
  {Vandevondele}}, \bibinfo {author} {\bibfnamefont {U.}~\bibnamefont
  {Borstnik}}, \ and\ \bibinfo {author} {\bibfnamefont {J.}~\bibnamefont
  {Hutter}},\ }\href@noop {} {\bibfield  {journal} {\bibinfo  {journal}
  {Journal Of Chemical Theory And Computation}\ }\textbf {\bibinfo {volume}
  {8}},\ \bibinfo {pages} {3565} (\bibinfo {year} {2012})}\BibitemShut
  {NoStop}%
\bibitem [{\citenamefont {Kotliar}\ \emph {et~al.}(2006)\citenamefont
  {Kotliar}, \citenamefont {Savrasov}, \citenamefont {Haule}, \citenamefont
  {Oudovenko}, \citenamefont {Parcollet},\ and\ \citenamefont
  {Marianetti}}]{Kotliar2006865}%
  \BibitemOpen
  \bibfield  {author} {\bibinfo {author} {\bibfnamefont {G.}~\bibnamefont
  {Kotliar}}, \bibinfo {author} {\bibfnamefont {S.~Y.}\ \bibnamefont
  {Savrasov}}, \bibinfo {author} {\bibfnamefont {K.}~\bibnamefont {Haule}},
  \bibinfo {author} {\bibfnamefont {V.~S.}\ \bibnamefont {Oudovenko}}, \bibinfo
  {author} {\bibfnamefont {O.}~\bibnamefont {Parcollet}}, \ and\ \bibinfo
  {author} {\bibfnamefont {C.~A.}\ \bibnamefont {Marianetti}},\ }\href@noop {}
  {\bibfield  {journal} {\bibinfo  {journal} {Rev. Mod. Phys.}\ }\textbf
  {\bibinfo {volume} {78}},\ \bibinfo {pages} {865} (\bibinfo {year}
  {2006})}\BibitemShut {NoStop}%
\bibitem [{\citenamefont {Vanderbilt}\ \emph {et~al.}(1989)\citenamefont
  {Vanderbilt}, \citenamefont {Taole},\ and\ \citenamefont
  {Narasimhan}}]{Vanderbilt19895657}%
  \BibitemOpen
  \bibfield  {author} {\bibinfo {author} {\bibfnamefont {D.}~\bibnamefont
  {Vanderbilt}}, \bibinfo {author} {\bibfnamefont {S.~H.}\ \bibnamefont
  {Taole}}, \ and\ \bibinfo {author} {\bibfnamefont {S.}~\bibnamefont
  {Narasimhan}},\ }\href@noop {} {\bibfield  {journal} {\bibinfo  {journal}
  {Phys. Rev. B}\ }\textbf {\bibinfo {volume} {40}},\ \bibinfo {pages} {5657}
  (\bibinfo {year} {1989})}\BibitemShut {NoStop}%
\bibitem [{\citenamefont {Kingsmith}\ and\ \citenamefont
  {Vanderbilt}(1994)}]{Kingsmith19945828}%
  \BibitemOpen
  \bibfield  {author} {\bibinfo {author} {\bibfnamefont {R.~D.}\ \bibnamefont
  {Kingsmith}}\ and\ \bibinfo {author} {\bibfnamefont {D.}~\bibnamefont
  {Vanderbilt}},\ }\href@noop {} {\bibfield  {journal} {\bibinfo  {journal}
  {Phys. Rev. B}\ }\textbf {\bibinfo {volume} {49}},\ \bibinfo {pages} {5828}
  (\bibinfo {year} {1994})}\BibitemShut {NoStop}%
\bibitem [{\citenamefont {Zhong}\ \emph {et~al.}(1994)\citenamefont {Zhong},
  \citenamefont {Vanderbilt},\ and\ \citenamefont {Rabe}}]{Zhong19941861}%
  \BibitemOpen
  \bibfield  {author} {\bibinfo {author} {\bibfnamefont {W.}~\bibnamefont
  {Zhong}}, \bibinfo {author} {\bibfnamefont {D.}~\bibnamefont {Vanderbilt}}, \
  and\ \bibinfo {author} {\bibfnamefont {K.~M.}\ \bibnamefont {Rabe}},\
  }\href@noop {} {\bibfield  {journal} {\bibinfo  {journal} {Phys. Rev. Lett.}\
  }\textbf {\bibinfo {volume} {73}},\ \bibinfo {pages} {1861} (\bibinfo {year}
  {1994})}\BibitemShut {NoStop}%
\bibitem [{\citenamefont {Zhong}\ \emph {et~al.}(1995)\citenamefont {Zhong},
  \citenamefont {Vanderbilt},\ and\ \citenamefont {Rabe}}]{Zhong19956301}%
  \BibitemOpen
  \bibfield  {author} {\bibinfo {author} {\bibfnamefont {W.}~\bibnamefont
  {Zhong}}, \bibinfo {author} {\bibfnamefont {D.}~\bibnamefont {Vanderbilt}}, \
  and\ \bibinfo {author} {\bibfnamefont {K.~M.}\ \bibnamefont {Rabe}},\
  }\href@noop {} {\bibfield  {journal} {\bibinfo  {journal} {Phys. Rev. B}\
  }\textbf {\bibinfo {volume} {52}},\ \bibinfo {pages} {6301} (\bibinfo {year}
  {1995})}\BibitemShut {NoStop}%
\bibitem [{\citenamefont {Keating}(1966{\natexlab{a}})}]{Keating1966637}%
  \BibitemOpen
  \bibfield  {author} {\bibinfo {author} {\bibfnamefont {P.~N.}\ \bibnamefont
  {Keating}},\ }\href@noop {} {\bibfield  {journal} {\bibinfo  {journal}
  {Physical Review}\ }\textbf {\bibinfo {volume} {145}},\ \bibinfo {pages}
  {637} (\bibinfo {year} {1966}{\natexlab{a}})}\BibitemShut {NoStop}%
\bibitem [{\citenamefont {Keating}(1966{\natexlab{b}})}]{Keating1966674}%
  \BibitemOpen
  \bibfield  {author} {\bibinfo {author} {\bibfnamefont {P.~N.}\ \bibnamefont
  {Keating}},\ }\href@noop {} {\bibfield  {journal} {\bibinfo  {journal}
  {Physical Review}\ }\textbf {\bibinfo {volume} {149}},\ \bibinfo {pages}
  {674} (\bibinfo {year} {1966}{\natexlab{b}})}\BibitemShut {NoStop}%
\bibitem [{\citenamefont {Pytte}(1972)}]{Pytte19723758}%
  \BibitemOpen
  \bibfield  {author} {\bibinfo {author} {\bibfnamefont {E.}~\bibnamefont
  {Pytte}},\ }\href@noop {} {\bibfield  {journal} {\bibinfo  {journal} {Phys.
  Rev. B}\ }\textbf {\bibinfo {volume} {5}},\ \bibinfo {pages} {3758} (\bibinfo
  {year} {1972})}\BibitemShut {NoStop}%
\bibitem [{\citenamefont {Rabe}\ and\ \citenamefont
  {Waghmare}(1995)}]{Rabe199513236}%
  \BibitemOpen
  \bibfield  {author} {\bibinfo {author} {\bibfnamefont {K.~M.}\ \bibnamefont
  {Rabe}}\ and\ \bibinfo {author} {\bibfnamefont {U.~V.}\ \bibnamefont
  {Waghmare}},\ }\href@noop {} {\bibfield  {journal} {\bibinfo  {journal}
  {Phys. Rev. B}\ }\textbf {\bibinfo {volume} {52}},\ \bibinfo {pages} {13236}
  (\bibinfo {year} {1995})}\BibitemShut {NoStop}%
\bibitem [{\citenamefont {Iniguez}\ \emph {et~al.}(2000)\citenamefont
  {Iniguez}, \citenamefont {Garcia},\ and\ \citenamefont
  {Perez-mato}}]{Iniguez20003127}%
  \BibitemOpen
  \bibfield  {author} {\bibinfo {author} {\bibfnamefont {J.}~\bibnamefont
  {Iniguez}}, \bibinfo {author} {\bibfnamefont {A.}~\bibnamefont {Garcia}}, \
  and\ \bibinfo {author} {\bibfnamefont {J.~M.}\ \bibnamefont {Perez-mato}},\
  }\href@noop {} {\bibfield  {journal} {\bibinfo  {journal} {Physical Review
  B}\ }\textbf {\bibinfo {volume} {61}},\ \bibinfo {pages} {3127} (\bibinfo
  {year} {2000})}\BibitemShut {NoStop}%
\bibitem [{\citenamefont {Esfarjani}\ and\ \citenamefont
  {Stokes}(2008)}]{Esfarjani2008144112}%
  \BibitemOpen
  \bibfield  {author} {\bibinfo {author} {\bibfnamefont {K.}~\bibnamefont
  {Esfarjani}}\ and\ \bibinfo {author} {\bibfnamefont {H.~T.}\ \bibnamefont
  {Stokes}},\ }\href {\doibase 10.1103/PhysRevB.77.144112} {\bibfield
  {journal} {\bibinfo  {journal} {Phys. Rev. B}\ }\textbf {\bibinfo {volume}
  {77}},\ \bibinfo {pages} {144112} (\bibinfo {year} {2008})}\BibitemShut
  {NoStop}%
\bibitem [{\citenamefont {Esfarjani}\ \emph {et~al.}(2011)\citenamefont
  {Esfarjani}, \citenamefont {Chen},\ and\ \citenamefont
  {Stokes}}]{PhysRevB.84.085204}%
  \BibitemOpen
  \bibfield  {author} {\bibinfo {author} {\bibfnamefont {K.}~\bibnamefont
  {Esfarjani}}, \bibinfo {author} {\bibfnamefont {G.}~\bibnamefont {Chen}}, \
  and\ \bibinfo {author} {\bibfnamefont {H.~T.}\ \bibnamefont {Stokes}},\
  }\href {\doibase 10.1103/PhysRevB.84.085204} {\bibfield  {journal} {\bibinfo
  {journal} {Phys. Rev. B}\ }\textbf {\bibinfo {volume} {84}},\ \bibinfo
  {pages} {085204} (\bibinfo {year} {2011})}\BibitemShut {NoStop}%
\bibitem [{\citenamefont {Shiomi}\ \emph {et~al.}(2011)\citenamefont {Shiomi},
  \citenamefont {Esfarjani},\ and\ \citenamefont {Chen}}]{PhysRevB.84.104302}%
  \BibitemOpen
  \bibfield  {author} {\bibinfo {author} {\bibfnamefont {J.}~\bibnamefont
  {Shiomi}}, \bibinfo {author} {\bibfnamefont {K.}~\bibnamefont {Esfarjani}}, \
  and\ \bibinfo {author} {\bibfnamefont {G.}~\bibnamefont {Chen}},\ }\href
  {\doibase 10.1103/PhysRevB.84.104302} {\bibfield  {journal} {\bibinfo
  {journal} {Phys. Rev. B}\ }\textbf {\bibinfo {volume} {84}},\ \bibinfo
  {pages} {104302} (\bibinfo {year} {2011})}\BibitemShut {NoStop}%
\bibitem [{\citenamefont {Shiga}\ \emph {et~al.}(2012)\citenamefont {Shiga},
  \citenamefont {Shiomi}, \citenamefont {Ma}, \citenamefont {Delaire},
  \citenamefont {Radzynski}, \citenamefont {Lusakowski}, \citenamefont
  {Esfarjani},\ and\ \citenamefont {Chen}}]{PhysRevB.85.155203}%
  \BibitemOpen
  \bibfield  {author} {\bibinfo {author} {\bibfnamefont {T.}~\bibnamefont
  {Shiga}}, \bibinfo {author} {\bibfnamefont {J.}~\bibnamefont {Shiomi}},
  \bibinfo {author} {\bibfnamefont {J.}~\bibnamefont {Ma}}, \bibinfo {author}
  {\bibfnamefont {O.}~\bibnamefont {Delaire}}, \bibinfo {author} {\bibfnamefont
  {T.}~\bibnamefont {Radzynski}}, \bibinfo {author} {\bibfnamefont
  {A.}~\bibnamefont {Lusakowski}}, \bibinfo {author} {\bibfnamefont
  {K.}~\bibnamefont {Esfarjani}}, \ and\ \bibinfo {author} {\bibfnamefont
  {G.}~\bibnamefont {Chen}},\ }\href {\doibase 10.1103/PhysRevB.85.155203}
  {\bibfield  {journal} {\bibinfo  {journal} {Phys. Rev. B}\ }\textbf {\bibinfo
  {volume} {85}},\ \bibinfo {pages} {155203} (\bibinfo {year}
  {2012})}\BibitemShut {NoStop}%
\bibitem [{\citenamefont {Wojdel}\ \emph {et~al.}(2013)\citenamefont {Wojdel},
  \citenamefont {Hermet}, \citenamefont {Ljungberg}, \citenamefont {Ghosez},\
  and\ \citenamefont {Iniguez}}]{Wojdel2013305401}%
  \BibitemOpen
  \bibfield  {author} {\bibinfo {author} {\bibfnamefont {J.~C.}\ \bibnamefont
  {Wojdel}}, \bibinfo {author} {\bibfnamefont {P.}~\bibnamefont {Hermet}},
  \bibinfo {author} {\bibfnamefont {M.~P.}\ \bibnamefont {Ljungberg}}, \bibinfo
  {author} {\bibfnamefont {P.}~\bibnamefont {Ghosez}}, \ and\ \bibinfo {author}
  {\bibfnamefont {J.}~\bibnamefont {Iniguez}},\ }\href@noop {} {\bibfield
  {journal} {\bibinfo  {journal} {Journal Of Physics-condensed Matter}\
  }\textbf {\bibinfo {volume} {25}},\ \bibinfo {pages} {305401} (\bibinfo
  {year} {2013})}\BibitemShut {NoStop}%
\bibitem [{\citenamefont {Behler}\ and\ \citenamefont
  {Parrinello}(2007)}]{Behler2007146401}%
  \BibitemOpen
  \bibfield  {author} {\bibinfo {author} {\bibfnamefont {J.}~\bibnamefont
  {Behler}}\ and\ \bibinfo {author} {\bibfnamefont {M.}~\bibnamefont
  {Parrinello}},\ }\href@noop {} {\bibfield  {journal} {\bibinfo  {journal}
  {Phys. Rev. Lett.}\ }\textbf {\bibinfo {volume} {98}},\ \bibinfo {pages}
  {146401} (\bibinfo {year} {2007})}\BibitemShut {NoStop}%
\bibitem [{\citenamefont {Eshet}\ \emph {et~al.}(2010)\citenamefont {Eshet},
  \citenamefont {Khaliullin}, \citenamefont {Kuhne}, \citenamefont {Behler},\
  and\ \citenamefont {Parrinello}}]{Eshet2010184107}%
  \BibitemOpen
  \bibfield  {author} {\bibinfo {author} {\bibfnamefont {H.}~\bibnamefont
  {Eshet}}, \bibinfo {author} {\bibfnamefont {R.~Z.}\ \bibnamefont
  {Khaliullin}}, \bibinfo {author} {\bibfnamefont {T.~D.}\ \bibnamefont
  {Kuhne}}, \bibinfo {author} {\bibfnamefont {J.}~\bibnamefont {Behler}}, \
  and\ \bibinfo {author} {\bibfnamefont {M.}~\bibnamefont {Parrinello}},\
  }\href@noop {} {\bibfield  {journal} {\bibinfo  {journal} {Phys. Rev. B}\
  }\textbf {\bibinfo {volume} {81}},\ \bibinfo {pages} {184107} (\bibinfo
  {year} {2010})}\BibitemShut {NoStop}%
\bibitem [{\citenamefont {Eshet}\ \emph {et~al.}(2012)\citenamefont {Eshet},
  \citenamefont {Khaliullin}, \citenamefont {Kuhne}, \citenamefont {Behler},\
  and\ \citenamefont {Parrinello}}]{Eshet2012115701}%
  \BibitemOpen
  \bibfield  {author} {\bibinfo {author} {\bibfnamefont {H.}~\bibnamefont
  {Eshet}}, \bibinfo {author} {\bibfnamefont {R.~Z.}\ \bibnamefont
  {Khaliullin}}, \bibinfo {author} {\bibfnamefont {T.~D.}\ \bibnamefont
  {Kuhne}}, \bibinfo {author} {\bibfnamefont {J.}~\bibnamefont {Behler}}, \
  and\ \bibinfo {author} {\bibfnamefont {M.}~\bibnamefont {Parrinello}},\
  }\href@noop {} {\bibfield  {journal} {\bibinfo  {journal} {Phys. Rev. Lett.}\
  }\textbf {\bibinfo {volume} {108}},\ \bibinfo {pages} {115701} (\bibinfo
  {year} {2012})}\BibitemShut {NoStop}%
\bibitem [{\citenamefont {Khaliullin}\ \emph {et~al.}(2011)\citenamefont
  {Khaliullin}, \citenamefont {Eshet}, \citenamefont {Kuhne}, \citenamefont
  {Behler},\ and\ \citenamefont {Parrinello}}]{Khaliullin2011693}%
  \BibitemOpen
  \bibfield  {author} {\bibinfo {author} {\bibfnamefont {R.~Z.}\ \bibnamefont
  {Khaliullin}}, \bibinfo {author} {\bibfnamefont {H.}~\bibnamefont {Eshet}},
  \bibinfo {author} {\bibfnamefont {T.~D.}\ \bibnamefont {Kuhne}}, \bibinfo
  {author} {\bibfnamefont {J.}~\bibnamefont {Behler}}, \ and\ \bibinfo {author}
  {\bibfnamefont {M.}~\bibnamefont {Parrinello}},\ }\href@noop {} {\bibfield
  {journal} {\bibinfo  {journal} {Nature Materials}\ }\textbf {\bibinfo
  {volume} {10}},\ \bibinfo {pages} {693} (\bibinfo {year} {2011})}\BibitemShut
  {NoStop}%
\bibitem [{\citenamefont {Nelson}\ \emph {et~al.}(2013)\citenamefont {Nelson},
  \citenamefont {Hart}, \citenamefont {Zhou},\ and\ \citenamefont
  {Ozolins}}]{Nelson2013035125}%
  \BibitemOpen
  \bibfield  {author} {\bibinfo {author} {\bibfnamefont {L.~J.}\ \bibnamefont
  {Nelson}}, \bibinfo {author} {\bibfnamefont {G.}~\bibnamefont {Hart}},
  \bibinfo {author} {\bibfnamefont {F.}~\bibnamefont {Zhou}}, \ and\ \bibinfo
  {author} {\bibfnamefont {V.}~\bibnamefont {Ozolins}},\ }\href@noop {}
  {\bibfield  {journal} {\bibinfo  {journal} {Phys. Rev. B}\ }\textbf {\bibinfo
  {volume} {87}},\ \bibinfo {pages} {035125} (\bibinfo {year}
  {2013})}\BibitemShut {NoStop}%
\bibitem [{\citenamefont {Cornwell}(1997)}]{Cornwell}%
  \BibitemOpen
  \bibfield  {author} {\bibinfo {author} {\bibfnamefont {J.}~\bibnamefont
  {Cornwell}},\ }\href@noop {} {\emph {\bibinfo {title} {Group Theory in
  Physics}}}\ (\bibinfo  {publisher} {Academic Press},\ \bibinfo {address}
  {London},\ \bibinfo {year} {1997})\BibitemShut {NoStop}%
\bibitem [{\citenamefont {Tinkham}(1964)}]{Tinkham}%
  \BibitemOpen
  \bibfield  {author} {\bibinfo {author} {\bibfnamefont {M.}~\bibnamefont
  {Tinkham}},\ }\href@noop {} {\emph {\bibinfo {title} {Group Theory and
  Quantum Mechanics}}}\ (\bibinfo  {publisher} {Dover},\ \bibinfo {address}
  {Mineola, New York},\ \bibinfo {year} {1964})\BibitemShut {NoStop}%
\bibitem [{\citenamefont {Gonze}\ and\ \citenamefont
  {Vigneron}(1989)}]{Gonze198913120}%
  \BibitemOpen
  \bibfield  {author} {\bibinfo {author} {\bibfnamefont {X.}~\bibnamefont
  {Gonze}}\ and\ \bibinfo {author} {\bibfnamefont {J.~P.}\ \bibnamefont
  {Vigneron}},\ }\href@noop {} {\bibfield  {journal} {\bibinfo  {journal}
  {Phys. Rev. B}\ }\textbf {\bibinfo {volume} {39}},\ \bibinfo {pages} {13120}
  (\bibinfo {year} {1989})}\BibitemShut {NoStop}%
\bibitem [{\citenamefont {Debernardi}\ \emph {et~al.}(1995)\citenamefont
  {Debernardi}, \citenamefont {Baroni},\ and\ \citenamefont
  {Molinari}}]{Debernardi19951819}%
  \BibitemOpen
  \bibfield  {author} {\bibinfo {author} {\bibfnamefont {A.}~\bibnamefont
  {Debernardi}}, \bibinfo {author} {\bibfnamefont {S.}~\bibnamefont {Baroni}},
  \ and\ \bibinfo {author} {\bibfnamefont {E.}~\bibnamefont {Molinari}},\
  }\href@noop {} {\bibfield  {journal} {\bibinfo  {journal} {Phys. Rev. Lett.}\
  }\textbf {\bibinfo {volume} {75}},\ \bibinfo {pages} {1819} (\bibinfo {year}
  {1995})}\BibitemShut {NoStop}%
\bibitem [{\citenamefont {Deinzer}\ \emph {et~al.}(2003)\citenamefont
  {Deinzer}, \citenamefont {Birner},\ and\ \citenamefont
  {Strauch}}]{Deinzer2003144304}%
  \BibitemOpen
  \bibfield  {author} {\bibinfo {author} {\bibfnamefont {G.}~\bibnamefont
  {Deinzer}}, \bibinfo {author} {\bibfnamefont {G.}~\bibnamefont {Birner}}, \
  and\ \bibinfo {author} {\bibfnamefont {D.}~\bibnamefont {Strauch}},\
  }\href@noop {} {\bibfield  {journal} {\bibinfo  {journal} {Phys. Rev. B}\
  }\textbf {\bibinfo {volume} {67}},\ \bibinfo {pages} {144304} (\bibinfo
  {year} {2003})}\BibitemShut {NoStop}%
\bibitem [{\citenamefont {Martin}(2008)}]{Martin2008}%
  \BibitemOpen
  \bibfield  {author} {\bibinfo {author} {\bibfnamefont {R.~M.}\ \bibnamefont
  {Martin}},\ }\href@noop {} {\emph {\bibinfo {title} {Electronic Structure:
  Basic Theory and Practical Methods}}}\ (\bibinfo  {publisher} {Cambridge
  University Press},\ \bibinfo {address} {New York},\ \bibinfo {year}
  {2008})\BibitemShut {NoStop}%
\bibitem [{\citenamefont {Perdew}\ \emph {et~al.}(1996)\citenamefont {Perdew},
  \citenamefont {Burke},\ and\ \citenamefont
  {Ernzerhof}}]{PhysRevLett.77.3865}%
  \BibitemOpen
  \bibfield  {author} {\bibinfo {author} {\bibfnamefont {J.~P.}\ \bibnamefont
  {Perdew}}, \bibinfo {author} {\bibfnamefont {K.}~\bibnamefont {Burke}}, \
  and\ \bibinfo {author} {\bibfnamefont {M.}~\bibnamefont {Ernzerhof}},\ }\href
  {\doibase 10.1103/PhysRevLett.77.3865} {\bibfield  {journal} {\bibinfo
  {journal} {Phys. Rev. Lett.}\ }\textbf {\bibinfo {volume} {77}},\ \bibinfo
  {pages} {3865} (\bibinfo {year} {1996})}\BibitemShut {NoStop}%
\bibitem [{\citenamefont {Kresse}\ and\ \citenamefont
  {Hafner}(1993)}]{Kresse1993558}%
  \BibitemOpen
  \bibfield  {author} {\bibinfo {author} {\bibfnamefont {G.}~\bibnamefont
  {Kresse}}\ and\ \bibinfo {author} {\bibfnamefont {J.}~\bibnamefont
  {Hafner}},\ }\href@noop {} {\bibfield  {journal} {\bibinfo  {journal} {Phys.
  Rev. B}\ }\textbf {\bibinfo {volume} {47}},\ \bibinfo {pages} {558} (\bibinfo
  {year} {1993})}\BibitemShut {NoStop}%
\bibitem [{\citenamefont {Kresse}\ and\ \citenamefont
  {Hafner}(1994)}]{Kresse199414251}%
  \BibitemOpen
  \bibfield  {author} {\bibinfo {author} {\bibfnamefont {G.}~\bibnamefont
  {Kresse}}\ and\ \bibinfo {author} {\bibfnamefont {J.}~\bibnamefont
  {Hafner}},\ }\href@noop {} {\bibfield  {journal} {\bibinfo  {journal} {Phys.
  Rev. B}\ }\textbf {\bibinfo {volume} {49}},\ \bibinfo {pages} {14251}
  (\bibinfo {year} {1994})}\BibitemShut {NoStop}%
\bibitem [{\citenamefont {Kresse}\ and\ \citenamefont
  {Furthmuller}(1996{\natexlab{a}})}]{Kresse199615}%
  \BibitemOpen
  \bibfield  {author} {\bibinfo {author} {\bibfnamefont {G.}~\bibnamefont
  {Kresse}}\ and\ \bibinfo {author} {\bibfnamefont {J.}~\bibnamefont
  {Furthmuller}},\ }\href@noop {} {\bibfield  {journal} {\bibinfo  {journal}
  {Computational Materials Science}\ }\textbf {\bibinfo {volume} {6}},\
  \bibinfo {pages} {15} (\bibinfo {year} {1996}{\natexlab{a}})}\BibitemShut
  {NoStop}%
\bibitem [{\citenamefont {Kresse}\ and\ \citenamefont
  {Furthmuller}(1996{\natexlab{b}})}]{Kresse199611169}%
  \BibitemOpen
  \bibfield  {author} {\bibinfo {author} {\bibfnamefont {G.}~\bibnamefont
  {Kresse}}\ and\ \bibinfo {author} {\bibfnamefont {J.}~\bibnamefont
  {Furthmuller}},\ }\href@noop {} {\bibfield  {journal} {\bibinfo  {journal}
  {Phys. Rev. B}\ }\textbf {\bibinfo {volume} {54}},\ \bibinfo {pages} {11169}
  (\bibinfo {year} {1996}{\natexlab{b}})}\BibitemShut {NoStop}%
\bibitem [{\citenamefont {Kresse}\ and\ \citenamefont
  {Joubert}(1999)}]{Kresse19991758}%
  \BibitemOpen
  \bibfield  {author} {\bibinfo {author} {\bibfnamefont {G.}~\bibnamefont
  {Kresse}}\ and\ \bibinfo {author} {\bibfnamefont {D.}~\bibnamefont
  {Joubert}},\ }\href@noop {} {\bibfield  {journal} {\bibinfo  {journal} {Phys.
  Rev. B}\ }\textbf {\bibinfo {volume} {59}},\ \bibinfo {pages} {1758}
  (\bibinfo {year} {1999})}\BibitemShut {NoStop}%
\bibitem [{\citenamefont {Jensen}\ \emph {et~al.}(2012)\citenamefont {Jensen},
  \citenamefont {Bozin}, \citenamefont {Malliakas}, \citenamefont {Stone},
  \citenamefont {Lumsden}, \citenamefont {Kanatzidis}, \citenamefont
  {Shapiro},\ and\ \citenamefont {Billinge}}]{Jensen2012085313}%
  \BibitemOpen
  \bibfield  {author} {\bibinfo {author} {\bibfnamefont {K.}~\bibnamefont
  {Jensen}}, \bibinfo {author} {\bibfnamefont {E.~S.}\ \bibnamefont {Bozin}},
  \bibinfo {author} {\bibfnamefont {C.~D.}\ \bibnamefont {Malliakas}}, \bibinfo
  {author} {\bibfnamefont {M.~B.}\ \bibnamefont {Stone}}, \bibinfo {author}
  {\bibfnamefont {M.~D.}\ \bibnamefont {Lumsden}}, \bibinfo {author}
  {\bibfnamefont {M.~G.}\ \bibnamefont {Kanatzidis}}, \bibinfo {author}
  {\bibfnamefont {S.~M.}\ \bibnamefont {Shapiro}}, \ and\ \bibinfo {author}
  {\bibfnamefont {S.}~\bibnamefont {Billinge}},\ }\href@noop {} {\bibfield
  {journal} {\bibinfo  {journal} {Phys. Rev. B}\ }\textbf {\bibinfo {volume}
  {86}},\ \bibinfo {pages} {085313} (\bibinfo {year} {2012})}\BibitemShut
  {NoStop}%
\bibitem [{\citenamefont {Delaire}\ \emph {et~al.}(2011)\citenamefont
  {Delaire}, \citenamefont {Ma}, \citenamefont {Marty}, \citenamefont {May},
  \citenamefont {Mcguire}, \citenamefont {Du}, \citenamefont {Singh},
  \citenamefont {Podlesnyak}, \citenamefont {Ehlers}, \citenamefont {Lumsden},\
  and\ \citenamefont {Sales}}]{Delaire2011614}%
  \BibitemOpen
  \bibfield  {author} {\bibinfo {author} {\bibfnamefont {O.}~\bibnamefont
  {Delaire}}, \bibinfo {author} {\bibfnamefont {J.}~\bibnamefont {Ma}},
  \bibinfo {author} {\bibfnamefont {K.}~\bibnamefont {Marty}}, \bibinfo
  {author} {\bibfnamefont {A.}~\bibnamefont {May}}, \bibinfo {author}
  {\bibfnamefont {M.}~\bibnamefont {Mcguire}}, \bibinfo {author} {\bibfnamefont
  {M.}~\bibnamefont {Du}}, \bibinfo {author} {\bibfnamefont {D.}~\bibnamefont
  {Singh}}, \bibinfo {author} {\bibfnamefont {A.}~\bibnamefont {Podlesnyak}},
  \bibinfo {author} {\bibfnamefont {G.}~\bibnamefont {Ehlers}}, \bibinfo
  {author} {\bibfnamefont {M.}~\bibnamefont {Lumsden}}, \ and\ \bibinfo
  {author} {\bibfnamefont {B.}~\bibnamefont {Sales}},\ }\href@noop {}
  {\bibfield  {journal} {\bibinfo  {journal} {Nature Materials}\ }\textbf
  {\bibinfo {volume} {10}},\ \bibinfo {pages} {614} (\bibinfo {year}
  {2011})}\BibitemShut {NoStop}%
\bibitem [{\citenamefont {Chen}\ \emph {et~al.}(2013)\citenamefont {Chen},
  \citenamefont {Ai},\ and\ \citenamefont {Marianetti}}]{Chen:2013}%
  \BibitemOpen
  \bibfield  {author} {\bibinfo {author} {\bibfnamefont {Y.}~\bibnamefont
  {Chen}}, \bibinfo {author} {\bibfnamefont {X.}~\bibnamefont {Ai}}, \ and\
  \bibinfo {author} {\bibfnamefont {C.~A.}\ \bibnamefont {Marianetti}},\
  }\href@noop {} {\bibfield  {journal} {\bibinfo  {journal} {arXiv:1312.6109
  [cond-mat.mtrl-sci]}\ } (\bibinfo {year} {2013})}\BibitemShut {NoStop}%
\end{thebibliography}%
